\documentclass[aps,twocolumn,showpacs]{revtex4-1}
\usepackage{graphicx}
\usepackage[all]{xy}
\usepackage{amsmath}
\usepackage{amssymb}
\usepackage{color}
\newcommand{\be}{\begin{equation}}
\newcommand{\ee}{\end{equation}}
\newcommand{\ben}{\begin{eqnarray}}
\newcommand{\een}{\end{eqnarray}}
\newcommand{\bes}{\begin{subequations}}
\newcommand{\ees}{\end{subequations}}

\begin{document}
\title{ $\delta-\delta^\prime$ generalized Robin boundary conditions \\ and \\ quantum vacuum fluctuations}

\author{J. M. Mu$\tilde{\rm n}$oz Casta$\tilde{\rm n}$eda,}
\affiliation{Institut f\"{u}r Theoretische Physik, Universit\"{a}t Leipzig, Germany}
\email{jose.munoz-castaneda@uni-leipzig.de}

\author{J. Mateos Guilarte,}
\affiliation{Departamento de F\'{\i}sica Fundamental, Universidad de Salamanca, Spain}
\email{guilarte@usal.es}

\date{\today}
\begin{abstract} 
The effects induced by the quantum vacuum fluctuations of one massless real scalar field on a configuration of two partially transparent plates 
are investigated. The physical properties of the infinitely thin plates are simulated by means of Dirac-$\delta-\delta^\prime$ point interactions.
It is shown that the distortion caused on the fluctuations by this external background gives rise to a generalization of Robin boundary conditions. The $T$-operator for potentials concentrated on points with non defined parity is evaluated with total generality. The quantum vacuum interaction energy between the two plates is computed in several dimensions using the $TGTG$ formula to find positive, negative, and zero Casimir energies. The parity properties of the $\delta-\delta^\prime$ potential demands to distinguish between opposite and identical objects. It is shown that between identical sets of $\delta-\delta^\prime$ plates repulsive, attractive, or null quantum vacuum forces arise. However there is always attraction between a pair of opposite $\delta-\delta^\prime$ plates.
\end{abstract}


\maketitle

\section{Introduction}
\label{sec:1}
More than thirty years ago  Symanzik in Ref. \cite{Symanzik:1981wd} established the correspondence between boundary conditions and surface interactions in quantum field theory. The most remarkable physical manifestation of surface interactions in QFT is the Casimir effect \cite{Casimir:1948dh}. Recently Bordag, Milton, Fosco and others have treated different idealized semitransparent plates by using Dirac-$\delta$ functions, see Refs. \cite{Fosco:2009zc,Parashar:2012it,Barton2004,Bordag:2005by,Bordag:2004dn,Fosco:2012gp,Milton:2013bm,Munoz-Castaneda:2013yga}. This idealization provides an analytic approach to study the electromagnetic quantum vacuum interaction between several types of material plates. In Reference \cite{Munoz-Castaneda:2013yga} we thoroughly studied the quantum vacuum interaction between two Dirac-$\delta$ plates with arbitrary couplings in one spatial dimension. If one or two of the Dirac-$\delta$ couplings are negative the quantum vacuum energy becomes imaginary and the phenomenon of fluctuation absorption/emission appears. A similar idea, based on 2D Dirac-$\delta$s, has been applied by Munoz-Castaneda and Bordag to analyse the quantum vacuum interaction between two cosmic strings, see Ref. \cite{munoz-bordag2014}. 

Quantum boundary conditions, compatible with the principles of QFT, that quantum fields can satisfy in a QFT defined over a domain with boundary has been studied by Munoz-Castaneda, Asorey {\it et al} during the past few years \cite{mc-asorey,MunozCastaneda:2011mv,jmmc-phd,Asorey:2008xt,Munoz-kk} in full generality. The analysis of quantum mechanical systems defined in bounded domains as self-adjoint extensions of the free particle Hamiltonian developed by Asorey-Ibort-Marmo (AIM) in Ref. \cite{aim} is the starting point of the work of Munoz-Castaneda, Asorey {\it et al}. These authors select the eigenfunctions of the self-adjoint Hamiltonians as the one particle states of a QFT defined on a compact manifold with boundary. By doing this they characterise all the boundary conditions compatible with unitarity that quantum fields confined in finite domains can satisfy. Moreover, calculations of the vacuum energy and other important magnitudes in the QFT as functions over the space of allowed boundary conditions were also achieved, see Refs. \cite{Munoz-kk,mc-asorey}. Although several authors suggested the connection between surface/point-like Dirac-$\delta$ interactions and a kind of generalized
Dirichlet boundary conditions, see \cite{Bordag:2005by,Parashar:2012it,Milton:2013bm,Fosco:2009zc}, we used the same strategy of introducing $\delta$-point interactions in Reference \cite{Munoz-Castaneda:2013yga} to make contact with the Asorey-Munoz-Castaneda formalism without complete success. More complex point-like/surface interactions are needed to implement other Asorey-Munoz-Castaneda quantum boundary conditions, e.g., of Robin type.

In this paper we propose to add Dirac-$\delta^\prime$ potentials to the same points where $\delta$ interactions were introduced in order to investigate which Asorey-Munoz-Castaneda boundary conditions can be reproduced by point-like potentials of the form $\mu\delta(x)+\lambda\delta^\prime(x)$. In the past 20 years there has been a lot of activity on how to define the derivative of the Dirac-$\delta$ as a quantum mechanical potential, see Refs. \cite{seba,albev-jpa2013,Kurasov1996297,Gadella20091310,Boya:1994ww} to find different approaches to the problem. The most successful way to define the $\delta^\prime$ is to introduce it using a regularization. The two most rigorous regularized definitions of the $\delta^\prime$ are the one introduced by Kurasov and Gadella {\it et al} in Refs. \cite{Kurasov1996297,Gadella20091310} and the regularised definitionintroduced by  Seba, Alveverio-Fassari, and others in Refs. \cite{seba,albev-jpa2013}. We remark that the regularised potentials through different approaches are not equivalent, but illuminating discussions of the 
distinct outcomes are offered in Refs. \cite{Gadella20091310,christ03-jpa36,toynog07-jpa40}.  Here we define the Dirac-$\delta^\prime$-potential following the proposal of Kurasov and Gadella {\it et al} in Refs. \cite{Kurasov1996297,Gadella20091310}. In their approach, the $\delta^\prime$ potential is regularized by also including a Dirac-$\delta$ interaction at the same point. By doing this, scale invariance associated to a pure 1D $\delta^\prime$ potential is broken. The combined pair of $\delta-\delta^\prime$ point interactions is defined as a self-adjoint extension
of the free particle Hamiltonian by imposing natural matching conditions to $\mu\delta(x)+\lambda\delta^\prime(x)$ on the eigenfunctions at the origin. 

Fulling in Ref. \cite{Fulling:2005js} succeeded in implementing Robin boundary conditions by means of a quantum graph vertex with a Dirac-$\delta$ attached, somehow characterized by the matching conditions for the $\delta^\prime$-potential given by Seba in \cite{seba}. 
In Ref. \cite{Fosco:2013wpa} Fosco {\it et al} explain how a superconducting circuit experiment formed by a coplanar wave-guide ended on a SQUID 
is described by one quantum scalar field in $(1+1)$-dimensions subjected to generalized Robin boundary conditions at the endpoints of an interval. Our strategy, however, will be to consider two pairs of $\mu\delta(x)+\lambda\delta^\prime(x)$ interactions and study the quantum  fluctuations of a quantum scalar field in the ${\mathbb R}\times [-a,a]$ space-time under the influence of the static background: $U(x)=\mu_1\delta(x+a)+\lambda_1\delta^\prime(x+a)+\mu_2\delta(x-a)+\lambda_2\delta^\prime(x-a)$. We will follow and generalise the procedures and techniques
established in \cite{Munoz-Castaneda:2013yga} when the $\delta^\prime$s are switched off. Choosing the matching conditions of Gadella {\it et al}
we end in a richer situation where generalized Robin conditions arise linking this pair of point interactions to
quantum boundary conditions in QFT compatible with unitarity as described in the Asorey-Munoz-Castaneda formalism. Additionally interpretation 
of the point interactions as featuring two pairs of Casimir plates we will apply the $TGTG$ formula to evaluate the quantum vacuum energy. As we will demonstrate, in the $\mu_1>0, \mu_2>0$ cases, the vacuum energy between two $\delta-\delta^\prime$ plates  is found to be positive, negative, or zero giving rise respectively to repulsive, attractive, or null Casimir forces.

Whereas the couplings to the $\delta$ potentials physically describe the plasma frequencies in Barton's hydrodynamical model \cite{Barton2004} characterizing the electromagnetic properties of the conducting plates, the physical meaning of the coupling to the $\delta^\prime$-potential has been discovered only very recently. M. Bordag's analysis of monoatomically thin polarizable plates formed by lattices of dipoles published recently in \cite{bordag14-prd89} shows that $\delta^\prime$-potentials appear in the interaction between the electric field and the component of the diagonal polarizability tensor acting on the direction orthogonal to the plate, see sections I and II as well as equation (33) in Ref. \cite{bordag14-prd89}. The $\delta^\prime$ coupling $\lambda$ thus describes the response of the orthogonal polarizability of a monoatomically thin plate to the electromagnetic field. The Bordag paper paves the way to a finer understanding of the electromagnetic response of the monoatomically thin plate that completes previous works by G. Barton \cite{barton05-jpa,barton13-njp}, where the plasma model is used, and K. Milton and collaborators \cite{Parashar:2012it,milton13-ncc}, where the orthogonal polarizability was not accounted for.

Starting with a lightning review of the quantum mechanical spectrum of the $\delta-\delta^\prime$ point-potential defined by Kurasov-Gadella, we discuss as a novelty new physical properties of the $\delta-\delta^\prime$ and  the spectrum of the double $\delta-\delta^\prime$ potential. For the double $\delta-\delta^\prime$ potential the scattering amplitudes and the bound state energies,
identified graphically, and eigenfunctions are unveiled. The analysis is performed with reference to the space of parameters, i.e., the four couplings to the two pairs of point interaction, because it is important to know where bound states inducing absorption/emission phenomena arise: in these zones of the parameter space the scalar quantum field theory is not unitary. A second novel point is the demonstration that $\delta-\delta^\prime$ interactions provide a dynamical materialization of generalized Robin boundary conditions. The third achievement is the calculation of the quantum vacuum interaction energy between two $\delta-\delta^\prime$ plates by means of the $TGTG$ formalism. Numerical integrations in the $TGTG$ formula show that the Casimir forces between two of such plates can be attractive, repulsive, or null depending on
the chosen zone of the parameter space. In particular, several planes in the 4D parameter space will be chosen to present results offering several tomographic views of the problem. All this material will be organised as follows. In Section \ref{sec:2} the basic formulas are compiled. In particular, in Subsection \ref{ss:d-dp} we describe the quantum physics of the single $\delta-\delta^\prime$ potential. In Section \ref{sec:4} the quantum mechanical spectrum of two $\delta-\delta^\prime$ interactions is studied: we solve the scattering problem and characterise the bound state spectrum. In Section \ref{sec:5} we study the scalar quantum field theory that arises in the background of a double $\delta-\delta^\prime$ potential and the connection with Asorey-Munoz-Castaneda formalism for quantum fields in bounded domains via the relativistic probability flux. The Asorey-Munoz-Castaneda formalism is used in Section \ref{sec-robin} to demonstrate that the $\delta-\delta^\prime$ potential is a semitransparent generalization of Robin boundary conditions, and how the usual Robin boundary conditions arise in this background. In Section \ref{sec:6} the scalar quantum vacuum interaction between two $\delta-\delta^\prime$ plates is analyzed through the general formula for the $T$-operator of a point potential with non defined parity previously obtained. Finally, in Section \ref{conclusions} we discuss the results and draw the main conclusions.

\section{Quantum fluctuations of $1+1$-dimensional scalar fields}
\label{sec:2}

\subsection{The field equation and the Green function}
The fluctuations of 1D scalar fields on static classical backgrounds modelled by the function $U(x)$ are governed by the action:
\be\label{action1d}
S[\Phi]=\int\, d^2x \,\left[\frac{1}{2}\partial_\mu\Phi\partial^\mu\Phi-\frac{1}{2}U(x)\Phi^2(x,t)\right]
\ee
We shall focus on compact support functions $U(x)$ in order to deal with well-defined scattering problems \cite{galindo1990quantum}.
The classical field and the Green's function equations arising from (\ref{action1d}) are respectively
\ben
 &&\left(\partial_t^2-\partial_x^2+U(x)\right)\Phi(x,t)=0\\
 && \left(\partial_t^2-\partial_x^2+U(x)\right)G(x,t;x^\prime,t^\prime)=\delta(x-x^\prime)\delta(t-t^\prime).\nonumber
\een
 Performing a Fourier decomposition in the time coordinate of the scalar field
 \be
 \Phi(t,x)=\int_{-\infty}^\infty \, \frac{d\omega}{2\pi}\, e^{i\omega t}\phi_\omega(x),
  \ee
the general solution of the field equation becomes a linear superposition of the eigenfunctions of the static fluctuation Schr\"odinger operator:
 \be
 -\phi^{\prime\prime}_\omega(x)+U(x)\phi_\omega(x)=\omega^2 \phi_\omega(x).
 \ee
The same Fourier decomposition leads to the reduced Green's function $G_\omega(x,x^\prime)$ and its corresponding differential equation:
\ben\label{red-green}
&& G(x,t;x^\prime,t^\prime)=\int_{-\infty}^\infty \, \frac{d\omega}{2\pi}\, e^{i\omega(t^\prime -t)}\, G_\omega(x,x')\\
&&\left(-\omega^2-d^2/dx^2+U(x)\right)G_\omega(x,x')=\delta(x-x').
\een
The reduced Green function plays a central r$\hat{\rm o}$le in the paper. We shall need the reduced Green function in the calculation of the Casimir energy by means of the $TGTG$ formalism developed in References \cite{Bordag:2009zz,Kenneth:2007jk,Milton:2004ya,Bordag:2011aa}.

\subsection{One-particle scattering waves and bound states}

The one-particle states of the $(1+1)D$ scalar quantum field theory are the eigenfunctions of the Schr\"odinger operator
\begin{equation}
  {\bf K}={\bf K}^0+U(x)=-\frac{d^2}{dx^2}+U(x) \, \,  .
\end{equation}
Generically this operator has both continuous  and discrete spectrum
\begin{eqnarray}
  \hspace{-0.8cm}{\bf K}\psi_j(x)&=&\omega_j^2\psi_j(x),\quad j=1,2,...,l, \,\,l\in\mathbb{N}\label{disck}\\
  \hspace{-0.8cm}{\bf K}\psi_k(x)&=&\omega(k)^2\psi_k(x;k),\,\,\omega(k)^2=k^2 \,\, , k\in\mathbb{R}\label{scattk}
\end{eqnarray}
For each $k\in\mathbb{R}$ the differential equation (\ref{scattk}) has two linear independent solutions: scattering waves incoming from the left $\psi_k^{(R)}(x)$ and from the right $\psi_k^{(L)}(x)$. Their asymptotic behaviour is determined by the scattering amplitudes, see e.g. References \citep{2008Boya,galindo1990quantum}:
\ben\label{asympscatt}
&&\psi_k^{(R)}(x)\simeq\left\{\begin{tabular}{ccc}
$e^{ikx}+r_{_R}(k)e^{-ikx}$ & $,$ & $x\to -\infty$ \\
$t(k)e^{ikx}$ & $,$ & $x\to \infty$ \\
\end{tabular}
\right.\label{asympscattR}\\
&&\psi_k^{(L)}(x)\simeq\left\{\begin{tabular}{ccc}
$t(k)e^{-ikx}$& $,$ & $x\to -\infty$ \\
$e^{-ikx}+r_{_L}(k)e^{ikx} $& $,$ & $x\to \infty$ \\
\end{tabular}
\right. .\label{asympscattL}
\een
The Wronskian of the two independent scattering solutions is proportional to the transmission amplitude $t(k)$
\begin{equation}
  W[\psi_k^{(R)}(x),\psi_k^{(L)}(x)]=-2ik\, t(k)\equiv W_{RL}(k)\, \, ,
\end{equation}
which is identical for $\psi^{(R)}$ and $\psi^{(L)}$ waves due to time-reversal invariance.
The reduced Green function defined in (\ref{red-green}) is obtained from the two independent scattering solutions by means of the following expression, see e.g. \citep{Bordag:2009zz}:
\ben
G_\omega(x,x^\prime)&=&\frac{1}{W_{RL}(k)}\left(\theta(x-x^\prime)\psi_k^{(R)}(x)\psi_k^{(L)}(x^\prime)\right.\nonumber\\&&
\left.+\theta(x^\prime-x)\psi_k^{(R)}(x^\prime)\psi_k^{(L)}(x)\right),\label{red-green-scatt}
\een
where $\theta(x)$ is the Heaviside step function.

\subsection{The $TGTG$ method in $(1+1)$-dimensional theories}
In References \cite{Bordag:2009zz,Kenneth:2007jk,Milton:2004ya,Bordag:2011aa} one finds the description of the logical steps and equations leading to the $TGTG$ formula for the Casimir energy/quantum vacuum interaction between two compact/topological disjoint objects in $(1+1)$-dimensional scalar quantum field theories. We offer here a brief summary. The Lipmann-Schwinger equation arising in quantum mechanical scattering theory defines the transfer matrix, also called $T$-operator, as
\begin{equation}\label{l-st}
  {\bf G}_\omega= {\bf G}_\omega^{(0)}-\frac{ {\bf G}_\omega^{(0)}\cdot{\bf U}\cdot{\bf G}_\omega^{(0)}}{{\bf I}+{\bf U}\cdot{\bf G}_\omega^{(0)}}\equiv {\bf G}_\omega^{(0)}\cdot({\bf I}-{\bf T}_\omega\cdot{\bf G}_\omega^{(0)})
\end{equation}
where ${\bf G}_\omega^{(0)}$ is the Green's function for the free particle operator ${\bf K}^0$, see again References \cite{2008Boya,galindo1990quantum}. It is convenient to write the last equality in equation (\ref{l-st}) in terms of the corresponding integral kernels:
\begin{eqnarray*}
  &&G_\omega(x,y)=G_\omega^{(0)}(x,y)-\\&&-\int dz_1dz_2
  G_\omega^{(0)}(x,z_1)T_\omega(z_1,z_2)G_\omega^{(0)}(z_2,y).
\end{eqnarray*}
The integral kernel of the $T$ operator in turns reads :
\begin{equation}
T_\omega(x,y)=U(x)\delta(x-y)+U(x)G_\omega^{(0)}(x,y)U(y)\, ,
\end{equation}
according to the detailed demonstration available in \cite{Bordag:2011aa}.

Compact disjoint objects in one dimension are modelled by potentials of the form
\begin{equation*}
  U(x)=U_1(x)+U_2(x),
\end{equation*}
where the smooth functions $U_i(x)$, $i=1,\,2$, have disjoint compact supports on the real line. Under this assumption the $TGTG$ formula for the vacuum interaction energy is \cite{Kenneth:2007jk}
\begin{equation}\label{tgtg-gen}
  E_0^{{\rm int}}=-\frac{i}{2}\int_0^\infty\frac{d\omega}{\pi}{\rm Tr}_{L^2}\ln\left({\bf 1}-{\bf M}_\omega\right) \, \, ,
\end{equation}
whereas the operator ${\bf M}_\omega$ and its integral kernel are:
\begin{eqnarray}
   {\bf M}_\omega&=& {\bf G}^{(0)}_\omega {\bf T}_\omega^{(1)}{\bf G}^{(0)}_\omega {\bf T}_\omega^{(2)}\label{defMop}\\
  M_\omega(x,y)&=&\nonumber{\int dz_1dz_2dz_3 \left[G^{(0)}_\omega(x,z_1)T_\omega^{(1)}(z_1,z_2)\times\right.} \\&\times&\left.G^{(0)}_\omega(z_2,z_3) T_\omega^{(2)}(z_3,y)\right]\, \, . \label{kerMgen}
\end{eqnarray}
Here, ${\bf T}^{(i)}_\omega$, $i=1,\,2$, is the $T$ operator associated to the object characterized by $U_i(x)$, $i=1,\,2$. The potentials $U_i(x)$, $i=1,\,2$, independently define two Schr\"odinger operators:
\begin{equation}
{\bf K}^{(i)}=-\frac{d^2}{dx^2}+U_i(x),\quad i=1,2.
\end{equation}
In general the operators ${\bf K}^{(i)}$ act on Hilbert spaces that are not isomorphic to the Hilbert space spanned by the eigenstates of the operator ${\bf K}^0$\footnote{When the ${\bf K}^{(i)}$ operator exhibits both discrete and continuous spectrum the Hilbert space spanned by the eigenstates of ${\bf K}^{(i)}$ is not isomorphic to the Hilbert space of plane waves where the operator ${\bf K}^0$ acts.}. The poles of the Green function ${\bf G}^{(0)}$ of the Klein-Gordon operator
\[
\Big(\partial_t^2-\partial_x^2\Big)G^{(0)}(x-x^\prime)=\delta(t-t^\prime)\delta(x-x^\prime)
\]
correspond to the propagation of the free mesons on shell. A Wick rotation $\omega\to i\xi$ in the energy-momentum plane skips all these poles 
and improves the convergence of the free quantum field theory. The same trick for the Green function in the external background does not work
if ${\bf K}^{(i)}$ presents bound states that still produce dangerous poles. In this case ${\bf G}_\omega^{(0)}$ and ${\bf T}^{(i)}_\omega$ refer to different Hilbert spaces. The product ${\bf G}^{(0)}_\omega\cdot{\bf T}^{(i)}_\omega$ is ill-defined and the formula (\ref{tgtg-gen}) is not valid. This problem is avoided in two steps: 1) Go to the Euclidean rotated quantum theory. 2) Push upwards the scattering threshold of the background distorted propagations by means of an infrared cutoff $m^2$ until the bound state eigenvalues disappear in the continuos spectrum. Having the bound states purely imaginary momenta of the form $k=i\kappa_b$, $\kappa_b>0$, the contributions of the bound states to the meson propagator
\begin{equation*}
\frac{i}{-\omega^2-\kappa_b^2+m^2} \, \, \Longrightarrow \, \, \frac{i}{\xi^2-\kappa_b^2+m^2} 
\end{equation*}
become finite provided that $m^2>\kappa_b^2$, $\forall\kappa_b$. After that, all the operators act over the same Hilbert space and the $TGTG$ formula reads
\begin{equation}\label{tgtg-gen-eucl}
  E_0^{{\rm int}}=\frac{1}{2}\int_0^\infty\frac{d\xi}{\pi}{\rm Tr}_{L^2}\ln\left({\bf 1}-{\bf M}_{i\xi}\right) \, ,
\end{equation}
where all the operators are the Euclidean rotated version of their Minkowskian counterparts and the dispersion relation energy-momentum incorporates the cutoff $m^2$, see Refs. \cite{Bordag:2011aa,Bordag:2009zz}. In this paper, however, we shall consider external backgrounds not allowing bound states in such a way that we keep the threshold at $m^2=0$.

\subsection{The $\delta-\delta'$ interaction.}\label{ss:d-dp}
We define the singular potential $V(x)=\mu\delta(x)+\lambda\delta^\prime(x)$ as the self-adjoint extension of the free Schr$\ddot{\rm o}$dinger operator on the real line excluding the origin built by Kurasov and  Gadella et al in References \cite{Kurasov1996297,Gadella20091310}. The potential depends on two real parameters $\mu$ and $\lambda$ that have dimensions of length to the $-1$ and $0$ respectively that sets the strength of the interactions. The spectral problem associated with the Hamiltonian
\begin{equation}
\widehat{\bf K}=-\frac{d^2}{dx^2}+\mu\delta(x)+\lambda\delta'(x) \label{ddpham}
\end{equation}
is defined as in \cite{Gadella20091310}. The matching conditions at $x=0$ are set to be {\footnote{In the Appendix it is shown how these matching conditions are tantamount to unitary matrices that correspond to boundary conditions acceptable in $1+1$ scalar QFT.}}:
\begin{eqnarray}
&&\psi_k(0\uparrow)=\frac{1+\lambda/2}{1-\lambda/2}\psi_k(0\downarrow) ;\label{1ddp-matchcond1}\\
&&\psi_k^\prime(0\uparrow)=\frac{1-\lambda/2}{1+\lambda/2}\psi_k^\prime(0\downarrow) +\frac{\mu}{1-\lambda^2/4}\psi_k(0\downarrow) .
\label{1ddp-matchcond2}
\end{eqnarray}
Here, and in the sequel, we denote as $f(a\uparrow)$ and $f(a\downarrow)$ respectively the limit of $f(x)$ at $x=a$ coming respectively from the right and from the left. Thus, the eigenwaves of a free quantum non relativistic particle
\begin{equation}
-\frac{d^2}{dx^2}\psi_k(x)=k^2\psi_k(x)
\end{equation}
are required to comply with the conditions (\ref{1ddp-matchcond1})-(\ref{1ddp-matchcond2}) in order to define the $\mu\delta(x)+\lambda\delta^\prime(x)$ interaction. There are two linearly independent scattering solutions if $k\in\mathbb{R}$: 1) incoming towards the origin from the far left $\psi_k^{(R)}$ plane waves and 2) incoming from the far right  $\psi_k^{(L)}$ plane waves. The effect of the interaction is encoded in the corresponding scattering amplitudes
\begin{eqnarray*}
\psi_k^{(R)}(x)&=&\left(e^{ikx}+r_R(k) e^{-ikx}\right)\theta(-x)+t_R(k) e^{ikx}\theta(x) \\ \psi_k^{(L)}(x)&=&t_L(k) e^{-ikx}\theta(-x) +\left(e^{-ikx}+r_L(k) e^{ikx}\right)\theta(x) \, ,
\end{eqnarray*}
where $\theta(x)$ denotes the Heaviside step function. Plugging these expressions in (\ref{1ddp-matchcond1}-\ref{1ddp-matchcond2}) one obtains two linear systems of two linear equations each that allow to obtain the following transmission and reflection coefficients
\begin{eqnarray}
&&t_R(k)=t_L(k)=t(k)=-\frac{k \left(\lambda ^2-4\right)}{k \left(\lambda ^2+4\right)+2 i \mu }\label{scata}\\
&&r_R(k)=\frac{-4 k \lambda -2 i \mu }{k \left(\lambda ^2+4\right)+2 i \mu };\,\, r_L(k)= \frac{4 k \lambda -2 i \mu }{k \left(\lambda ^2+4\right)+2 i \mu }\nonumber ,
\end{eqnarray}
in agreement with Ref. \cite{Gadella20091310}. Note that, because the Hamiltonian (\ref{ddpham}) defined this way is time-reversal invariant the right and left transmission amplitudes are equal. Parity invariance however is explicitly broken by the $\delta^\prime$ interaction. Therefore the right and left reflection amplitudes are different and related to the transmission amplitudes through the formula:
\begin{equation*}
t(k)-1=\frac{1}{2}\left(r_R(k)+r_L(k)\right) \, \, ,
\end{equation*}
in contrast to the pure $\delta$-potential the scattering is not unimodal. For $\lambda=\pm 2$ the transmission amplitude is zero; at the lines $(\mu,\pm 2)$ in the $\mu:\lambda$ plane the $\delta-\delta^\prime$ potential is completely opaque. The transition and reflection amplitudes at these lines become:
\begin{eqnarray}
&&\left.t(k)\right\vert_{\lambda=\pm 2}=0;\,\,\left.r_R\right\vert_{\lambda=2}=\left.r_L(k)\right\vert_{\lambda=-2}=-1 \nonumber\\
&&\left.r_R(k)\right\vert_{\lambda=-2}=\left.r_L(k)\right\vert_{\lambda=2}=\frac{4 k-i \mu }{4 k+i \mu } \, \, ,
\end{eqnarray}
a result also obtained but not underlined in \cite{Gadella20091310}. If the $\delta$-interaction is switched off, $\mu=0$, the scattering amplitudes become $k$-independent:
\begin{equation*}
\left. t\right\vert_{\mu=0}=\frac{4-\lambda ^2}{\lambda ^2+4}\, \, , \, \,
\left. r_R\right\vert_{\mu=0}=-\left. r_L\right\vert_{\mu=0}=-\frac{4 \lambda }{\lambda ^2+4} \, \, ,
\end{equation*}
as one expects from scale invariance. 

The transmission amplitude $t(k)$ has one pole over the imaginary axis at:
\begin{equation}
k_b=i\kappa_b \quad , \quad \kappa_b=-\frac{2 \mu }{\lambda ^2+4}.
\end{equation}
For $\mu<0$ this pole is due to the existence of a bound state with energy $E_b=-\kappa_b^2$, but when $\mu>0$ the pole corresponds to an anti-bound state. The bound/anti-bound state wave function reads:
\begin{eqnarray}
\psi_b(x)&=&\frac{(-\mu/2)^{1/2}}{1+\lambda^2/4}\left[(1+\lambda/2)e^{-\frac{\mu}{2(1+\lambda^2/4)}x}\theta(-x)\right.\nonumber\\
&+&\left.(1-\lambda/2)e^{\frac{\mu}{2(1+\lambda^2/4)}x}\theta(x)\right].
\end{eqnarray}
Clearly, when $\kappa_b<0$ the wave function $\psi_b$ is not normalizable but it becomes normalizable if $\kappa_b>0$. We remark that when $\lambda=-2$ there is probability density of finding the particle in this $\mu<0$ bound state only in the $x>0$ half-line 
but in the $\psi_b$ state the particle is located at the $x<0$ half-line if $\lambda=2$. The bound state completely disappears from the spectrum when $\mu=0$ because scale invariance.

\section{Two pairs of $\delta-\delta'$ interactions}
\label{sec:4}
Our main goal in this paper is the analytical description of the quantum vacuum interaction between two partially transparent plates in a Casimir setup by mimicking the plates as two point interactions, each of them of the form  $\mu\delta(x)+\lambda\delta'(x)$. The first task is to characterise the spectrum of the Schr$\ddot{\rm o}$dinger operator 
\begin{eqnarray}
\hat{\bf K}&=&-\frac{d^2}{dx^2}+\mu_1\delta(x+a)+\lambda_1\delta^\prime(x+a)\nonumber\\
&+&\mu_2 \delta(x-a)+\lambda_2\delta^\prime(x-a)\label{2ddp-ham}
\end{eqnarray}
in order to identify the eigenmodes of the scalar field fluctuations. The matching conditions between the one-particle wave functions at the
plate locations $x=\pm a$ generalize those defining a single $\delta-\delta^\prime$ interaction (\ref{1ddp-matchcond1})-(\ref{1ddp-matchcond2}):
\begin{widetext}
\begin{equation}
\left(\begin{array}{c}\vspace{0.12cm}\psi(-a\uparrow) \\ \vspace{0.12cm}\psi^\prime(-a\uparrow)\\ \vspace{0.12cm}\psi(a\uparrow)\\ \vspace{0.12cm}\psi^\prime(a\uparrow)\end{array}\right)= \left(\begin{array}{cccc}\frac{1+\lambda_1/2}{1-\lambda_1/2}
& 0 & 0 & 0 \\ \frac{\mu_1}{1-\lambda_1^2/4} & \frac{1-\lambda_1/2}{1+\lambda_1/2} & 0 & 0 \\ 0 & 0 & \frac{1+\lambda_2/2}{1-\lambda_2/2} & 0 \\ 0 & 0 & \frac{\mu_2}{1-\lambda_2^2/4} & \frac{1-\lambda_2/2}{1+\lambda_2/2} \end{array}\right)\left(\begin{array}{c}\vspace{0.12cm}\psi(-a\downarrow) \\ \vspace{0.12cm}\psi^\prime(-a\downarrow)\\ \vspace{0.12cm}\psi(a\downarrow)\\ \vspace{0.12cm}\psi^\prime(a\downarrow)\end{array}\right) \, \, .
\label{2ddp-matchcond}
\end{equation}
\end{widetext}
Thus, the plane waves $-\frac{d^2}{dx^2}\psi_k(x)=k^2\psi_k(x)$
are compelled to satisfy the matching conditions (\ref{2ddp-matchcond}). In order to have a detailed description of the spectrum of the Schr\"odinger operator (\ref{2ddp-ham}) we must study the scattering solutions and the bound states.
\subsection{Scattering waves}
Scattering states correspond to solutions with $k\in\mathbb{R}\Rightarrow k^2\geq 0$. The point interactions divide the real line into three zones: zone I $-a<x<a$, zone II $x<-a$, and zone III $x>a$. Given a value of $k\in \mathbb{R}$ there are two independent scattering solutions: 1)
incoming from the far left in zone II scattering waves $\psi_k^{(R)}$, 2) incoming from the far right in zone III scattering waves $\psi_k^{(L)}$. Away from the singular points the scattering states are of the form
\begin{equation}
\psi^{(R)}_k(x)=
\begin{cases}
 e^{-i k x}  r_R(k)+e^{i k x} & \, ,\,x\in \text{II} \\
 A_R(k) e^{i k x}+B_R(k) e^{-i k x} & \, , \, x\in \text{I} \\
 e^{i k x} t _R(k) & \, , \, x\in \text{III}
\end{cases},
\end{equation}
\vspace{-.5cm}
\begin{equation}
\psi^{(L)}_k(x)=\begin{cases}
 e^{-i k x} t_L(k) & \, , \, x\in \text{II} \\
 A_L(k) e^{i k x}+B_L(k) e^{-i k x} & \, , \,  x\in \text{I} \\
 e^{i k x} r_L(k)+e^{-i k x} & \, , \,  x\in \text{III}
\end{cases}.
\end{equation}
These solutions must satisfy the matching conditions (\ref{2ddp-matchcond}). Imposing (\ref{2ddp-matchcond}) over the two linear independent scattering states above gives rise to two linear systems of equations in $t$, $A$, $B$, $r$. Solving the linear systems we obtain $\{A_I,B_I,t_I,r_I\}_{I=R,L}$ as functions of $\{\lambda_1,\lambda_2,\mu_1,\mu_2,a,k\}$:
\begin{widetext}
\begin{eqnarray}
&&r_R(k)=\frac{-2}{\Delta(k)}\left( e^{2 i a k} \left(k \left(\lambda_1^2+4\right)-2 i\mu_1\right) (2 k\lambda_2+i\mu_2)+ e^{-2 i a k} (2 k \lambda_1+i \mu_1)
   \left(k \left(\lambda_2^2+4\right)+2 i \mu_2\right)\right)\label{rhor}\\
   &&A_R(k)= \frac{k (\lambda_1^2-4) (k (\lambda_2^2+4) + 2 i \mu_2)}{\Delta(k)}\,\, , \quad \, \, B_R(k)= -\frac{2 e^{2 i a k} k (-4 + \lambda_1^2) (2 k \lambda_2 + i \mu_2)}{\Delta(k)}\label{abr}\\
 &&r_L(k) =\frac{2}{\Delta(k)}\left(e^{-2 i a k} \left(k \left(\lambda_1^2+4\right)+2 i\mu_1\right) (2 k\lambda_2-i\mu_2)+^{2 i a k} (2 k \lambda_1-i \mu_1)
   \left(k \left(\lambda_2^2+4\right)-2 i \mu_2\right)\right)\label{rhol}\\
   && A_L(k)= -\frac{2e^{2iak}k(\lambda_2^2-4)(2k\lambda_1-i\mu_1)}{\Delta(k)}\,\, , \quad \, \, B_L(k)= -\frac{k (\lambda_2^2-4) (k (\lambda_1^2+4) + 2 i \mu_1)}{\Delta(k)}\label{abl}\\
   &&t_R(k)=t_L(k)=t(k)=\frac{(\lambda_1^2-4)(\lambda_2^2-4)k^2}{\Delta(k)}\label{sigma}\\
   &&\Delta(k)\equiv 4 e^{4 i a k} (2 k \lambda_1-i \mu_1) (2 k \lambda_2+i \mu_2)+\left(k \left(\lambda_1^2+4\right)+2 i \mu_1\right) \left(k
   \left(\lambda_2^2+4\right)+2 i \mu_2\right)\label{denom}
\end{eqnarray}
\end{widetext}
The coefficients $\{A_I,B_I,t_I,r_I\}_{I=R,L}$ completely determine the scattering states. The transmission amplitudes $t(k)$ are identical for $\psi_k^{(L)}$ and $\psi_k^{(R)}$ because of the time-reversal invariance meanwhile the reflection amplitudes $r_L(k)$, $r_R(k)$ are different because of the parity symmetry breaking. It is worthwhile to mention that the distance between singular points $2a$ explicitly breaks conformal invariance even if $\mu_1=\mu_2=0$. In this case
\begin{equation*}
\Delta(k)=k^2\Big(16\lambda_2\lambda_2e^{4a i k}+(\lambda_1^2+4)(\lambda_2^2+4) \Big)
\end{equation*}
and the scattering amplitudes depend on $k$.

\subsection{Bound states}
\label{sect-bound_st}

Poles of the transmission amplitude on the positive imaginary axis in the complex $k$-plane give rise to bound states of the Hamiltonian (\ref{2ddp-ham}). Thus the positive roots of the transcendent equation
\begin{equation*}
\Delta(i\kappa;\mu_1,\lambda_1,\mu_2,\lambda_2,a)=0 \quad , \quad \kappa\in\mathbb{R}^+
\end{equation*}
\begin{eqnarray}
&\Leftrightarrow &4 e^{-4  a \kappa} (2 \kappa \lambda_1- \mu_1) (2 \kappa \lambda_2+\mu_2)\nonumber\\&+&\left(\kappa \left(\lambda_1^2+4\right)+2 \mu_1\right) \left(\kappa\left(\lambda_2^2+4\right)+2  \mu_2\right)=0\label{boundst-eq}
\end{eqnarray}
are the bound states imaginary momenta. We define non-dimensional momenta $z\equiv\kappa a$ and coupling constants $\eta_i=\mu_i a \, \, , \, i=1,2$ and write equation (\ref{boundst-eq}) in the form:
\begin{equation}
4 e^{-4 z}=R(z;\eta_1,\eta_2,\lambda_1,\lambda_2) \label{rsboundst-eq} \, \, .
\end{equation}
The poles are the $z>0$ intersections between the exponential function in the left member of (\ref{rsboundst-eq}) and the rational function
of $z$ in the right member:
\begin{equation*}
R(z)\equiv -\frac{\left(z \left(\lambda_1^2+4\right)+2 \eta_1\right) \left(z\left(\lambda_2^2+4\right)+2  \eta_2\right)}{(2 z \lambda_1- \eta_1) (2 z \lambda_2+\eta_2)} \, \, \, .
\end{equation*}
The number of positive roots of (\ref{rsboundst-eq}) varies with the parameters $(\eta_1,\eta_2,\lambda_1,\lambda_2)$ and it is a quantity difficult to determine in full generality. Nevertheless, we list some properties of the function $R$ that will help in developing a qualitative analysis about the number of solutions on certain planes embedded in the four dimensional parameter space-

\begin{enumerate}

\item Limits: $\lim_{z\to 0} R(z;\eta_1,\eta_2,\lambda_1,\lambda_2)=4$
\begin{equation*}
\lim_{z\to +\infty} R(z;\eta_1,\eta_2,\lambda_1,\lambda_2)=-\frac{(4+\lambda_1^2)(4+\lambda_2^2)}{4\lambda_1\lambda_2}
\end{equation*}

\item The rational function $R(z;\eta_1,\eta_2,\lambda_1,\lambda_2)$ has two singular points: $z_1=\frac{\eta_1}{2\lambda_1}$ and $z_2=-\frac{\eta_2}{2\lambda_2}$. If $\lambda_1=-2$ $z_1$ becomes a regular point and the same fate happens to $z_2$ at $\lambda_2=2$.
\end{enumerate}
Together with this information, the knowledge of the value of the tangent to the curve $R(z)$ at the origin $R^\prime(0)$ with respect to the tangent of $4E^{-4 z}$ at $z=0$ will allow us to determine the number of bound states.

We now describe several reductions to a parameter subspace of two dimensions.

\paragraph{The double-$\delta$ interaction, or: $\lambda_1=\lambda_2=0$.}When the $\delta^\prime$ potentials are switched off the space of parameters reduces to the $\eta_1:\eta_2$ plane and there are zones of zero, one and two bound states separated by the two branches of the hyperbola: $2=-\frac{\eta_1+\eta_2}{\eta_1\eta_2}$, see Reference \cite{Munoz-Castaneda:2013yga}.

\paragraph{The double-$\delta^\prime$ system or: $\mu_1=\mu_2=0$.}The $\delta$ interactions are switched off and the transcendent equation (\ref{boundst-eq}) becomes:
\begin{equation}
e^{-4 z}=-\frac{(\lambda_2^2+4)(\lambda_1^2+4)}{16\lambda_1\lambda_2}\label{conf-boundst}
\end{equation}
The function of $\lambda_1,\lambda_2$ on the right member of this equation is a constant in $z$ such that it is either lesser than $-1$ if ${\rm sign}\lambda_1={\rm sign}\lambda_2$ or greater than $1$ if ${\rm sign}\lambda_1\neq {\rm sign}\lambda_2$. Therefore there are no intersections with the exponential in the left hand side of (\ref{conf-boundst}) for $z>0$ and hence there are no bound states. However there is an infinite tower of resonances. Writing $w=-iz$ we find the following complex solutions of (\ref{boundst-eq}): if $n\in{\mathbb Z}$ is an integer,
\begin{equation*}
w_n=k_n a=\frac{\pi}{2}(n+\frac{1}{2})-\frac{i}{4}\log\Big[\frac{(\lambda_2^2+4)(\lambda_1+4)^2}{16\lambda_1\lambda_2}\Big] \, \, , 
\end{equation*}
when ${\rm sign}\lambda_1={\rm sign}\lambda_2$, but 
\begin{equation*}
w_n=k_n a=\frac{\pi}{2}n-\frac{i}{4}\log\Big[\frac{(\lambda_2^2+4)(\lambda_1+4)^2}{16\vert\lambda_1\lambda_2\vert}\Big] \, \, , \, \, 
\end{equation*}
if ${\rm sign}\lambda_1\neq {\rm sign}\lambda_2$. These solutions have complex momenta with negative imaginary part. The real parts
come in pairs, $n+1/2$ with $-n-1/2$ ($n\geq 0$) if the signs of the couplings are equal, or $n$ with $-n$ ($n>0$) if the signs are different.
In this last case there exist one anti-bound state corresponding to $n=0$, the remaining solutions behaving as resonant states. It is a curious fact that the imaginary parts become null at the points $\lambda_1=\pm 2$, $\lambda=\pm 2$. In particular, if one of the couplings is $2$ and the other $-2$ the antibound state is a zero mode.

\begin{figure}[h]
\center{\includegraphics[width=8.25cm]{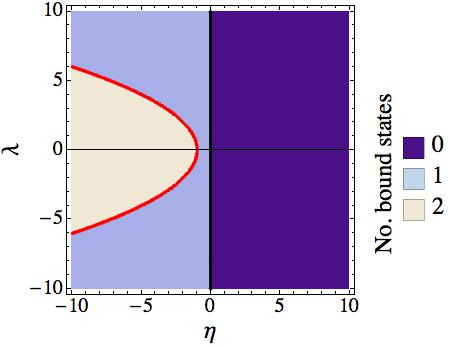}}
\caption{\footnotesize{Zones with different number of bound states in the $\lambda-\eta$ plane with the straight lines $\lambda=\pm 2$ excluded. The curve $R^\prime(0,\eta,\lambda)=-16$ (red line) divides the $\eta<0$ semi-plane into two zones: 1 bound state and two bound states.}}
\label{fig2}
\end{figure}

\paragraph{Two identical pairs of $\delta-\delta^\prime$ interactions: $\mu_1=\mu_2=\mu,\,\,\lambda_1=\lambda_2=\lambda$.}The rational function $R$ in the right member of the equation (\ref{boundst-eq}) reduces to
\begin{equation}
R(z,\eta,\lambda)=-\frac{(z(\lambda^2+4)+2\eta)^2}{(2\lambda z-\eta)(2\lambda z+\eta)}\, \, , \, \, \eta=\mu a ,
\end{equation}
The number of bound states bound characterized as the intersection of the two curves in the left and right members of the transcendent equation 
\begin{equation}
4 e^{-4z}=-\frac{(z(\lambda^2+4)+2\eta)^2}{(2\lambda z-\eta)(2\lambda z+\eta)}\label{bs-eqddp}
\end{equation}
is summarized as follows: If $\eta>0$ there are no bound states for any value of $\lambda$ including $\lambda=\pm 2$. If $\eta< 0$ is negative two situations are distinguished:
\begin{itemize}
\item $\eta<0$ and $\lambda\neq\pm2$. Two subcases arise in turn: when $0>R^\prime(0)>-16$ there are two bound states . If $-16>R^\prime(0)$ there is only bound state. The frontier between these two regimes is the curve $R^\prime(0)=-16$ in the half-plane $\eta<0$. In both cases, however, the
values of $\kappa_b$ corresponding to the intersections belong to the interval $(0,\vert\frac{\eta}{2\lambda}\vert)$.
In Fig \ref{fig2} it is shown how the separatrix curve divides the $\eta:\lambda$-plane into three zones: 1) zero bound states, 2) one bound state and 3) two bound states
\item $\eta<0$ and $\lambda=\pm2$. If $R^\prime(0)=32/\eta<-16$ the two curves do not intersect and there is no bound state.
If, alternatively, $-16<32/\eta<0$ there is one bound state. At $\lambda=\pm 2$ there exists one bound state only for $\eta<-2$, one bound state is lost with respect to $\lambda\neq \pm 2$ in each zone of the $\eta<0$ half-plane.
\end{itemize}

\begin{figure}[h]
\center{\includegraphics[width=8.25cm]{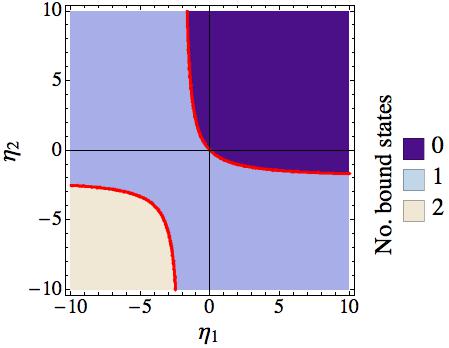}}
\caption{\footnotesize{Zones in the $\eta_1-\eta_2$ plane for $\lambda_1=2$ and $\lambda_2=-2$ with $0$, $1$, and $2$ bound states. The curve $R^\prime(0;\eta_1,\eta_2,2,-2)=-16$ (red line) divides the $\eta_1-\eta_2$ plane into three zones: no bound states, one bound state and two bound states.}}
\label{fig1}
\end{figure}

\paragraph{Two pairs of opaque interactions: $\lambda_1=\pm 2$, $\lambda_2=\pm 2$.}The previous result suggests the interest of considering the special values $\pm 2$ of the non dimensional couplings that produce opaque walls separately. Since the case of identical $\delta-\delta^\prime$ plates was previously analysed we consider only cases where $\eta_1\neq\eta_2$. Specifically, we distinguish four different possibilities:
\begin{itemize}
\item $\lambda_1=\lambda_2=2$. The bound state equation reads:
\begin{equation*}
e^{-4z}=\frac{ \eta _1+4 z}{\eta _1-4 z} \quad .
\end{equation*}
There is one solution to this equation if and only if $\eta_1<-2$, whereas $\eta_2$ can take any real value.

\item $\lambda_1=\lambda_2=-2$. The bound state equation becomes
\begin{equation*}
e^{-4z}=\frac{ \eta _2+4 z}{\eta _2-4 z}
\end{equation*}
and there is one solution to this equation if $\eta_2<-2$, whereas $\eta_1$ can take any real value

\item $\lambda_1=2,\,\,\lambda_2=-2$. In this case the bound state equation reads
\begin{equation}
e^{-4z}=\frac{ \left(\eta _1+4 z\right) \left(\eta _2+4 z\right)}{\left(\eta _1-4
   z\right) \left(\eta _2-4 z\right)} \label{bsddpop}
\end{equation}
 If $R^\prime >0$ there are no bound states, for $0<R^\prime(0)<-16$ there are two bound states, and there is only one if $-16<R^\prime(0)$. The zones with zero, one, and two bound states are plotted in Fig. \ref{fig1}.
\item $\lambda_1=-2,\,\,\lambda_2=2$. The bound state equation becomes $e^{-4z}=1\Leftrightarrow z=0$
and there are no bound states. 
\end{itemize}

To summarize we conclude that there are zones in the parameter space where $0$, $1$, or $2$ bound states exist in the spectrum of the Hamiltonian.
Resonances and anti-bound states do not belong to the spectrum because they are not normalizable. The exact values of bound state energies only
can be identified by graphical methods due to the transcendent character of the equations from which are determined. A novel fact with respect to
a similar analysis on the bound states of a double-$\delta$ system is that solutions of the transcendent equations for very large positive values of $\kappa$, i.e., very deep bound states, may exist if ${\rm sign}(\lambda_1)\neq {\rm sign}(\lambda_2)$ because in that case the function $R$ at $z=+\infty$ is positive.
\section{Scalar fields on compact spaces with boundary}
\label{sec:5}
On a cylindrical space-time ${\mathbb R}\times M$ where $M$ is a compact manifold of dimension $n$ with $n-1$-dimensional boundary $\partial M\equiv\Omega$ the action for a scalar field that generalizes (\ref{action1d}) reads:
\begin{eqnarray}
S(\phi)&=& \frac{1}{2}\int_{\mathbb R}\, dt \, \left[ \int_M \, d^nx \,\left\{\frac{\partial\phi^*}{\partial t}\cdot\frac{\partial\phi}{\partial t}-\right.\right.\nonumber \\  &-& \left. \left. \phi^*\Big(-\Delta+U(\vec{x})\Big)\phi\right\}-\int_\Omega\, d_\Omega \vec{\sigma}. \phi^*\vec{\nabla}\phi\right] \label{surfac} \, ,  
\end{eqnarray}
where $d_\Omega \vec{\sigma}$ is the volume differential element in the boundary time a unit vector poynting outwards $M$. The one-particle wave-functions $\phi(t_0,\vec{x})=\psi(\vec{x})$ belong to the space $L^2(M,\mathbb{C})$ of square integrable functions where the quantum mechanical 
Schr$\ddot{\rm o}$dinger operator
\begin{equation}
\hat{\bf K}=-\Delta + U(\vec{x}) \label{scho}
\end{equation}
acts symmetrically in general (see Ref. \cite{mc-asorey}).
\subsection{Probability flux conservation}
The Asorey-Munoz-Castaneda formalism to deal with quantum fields in bounded domains will be the basis of our approach, see the References \cite{Asorey:2008xt,jmmc-phd,mc-asorey,Munoz-kk}. Unitary scalar field theories on spaces with boundary are in one-to-one correspondence with the self-adjoint extensions of the $\hat{H}$ operator at the boundary. The continuity equation 
\begin{equation}
\frac{d}{d t}\int_M \rho =-\int_\Omega{\rm Re}(i\psi^*\overrightarrow{\nabla}\psi)\cdot d_\Omega\overrightarrow{\sigma}
\end{equation}
equals the variation in time of the probability with the flow of probability current across the boundary. Probability conservation, that is necessary for unitarity of the quantum field theory, demands annihilation of the flux at the boundary:
\begin{equation}
\int_\Omega{\rm Re}(i\psi^*\overrightarrow{\nabla}\psi)\cdot d_\Omega\overrightarrow{\sigma}=0 \label{probfl}
\end{equation}
Because the boundary conditions compatible with self-adjoint extensions of the Hamiltonian (\ref{scho}) may give rise to interferences between plane waves in the interior of $M$, the flux of the imaginary part of the current 
due to these interferences must also be null on the boundary for finding an unitary QFT. Therefore we require 
\begin{equation}
\Phi_\Omega(\psi)\equiv\int_\Omega(i\psi^*\overrightarrow{\nabla}\psi)\cdot d_\Omega\overrightarrow{\sigma}=0, \label{cprobfl}
\end{equation}
as explained in Refs. \cite{aim,mc-asorey}. The interpretation of the complex flux $\Phi_\Omega(\psi)$ is accordingly as follows:
\begin{itemize}
\item ${\rm Re}(\Phi_\Omega(\psi))$ is the probability flux across the surface $\Omega=\partial M$.
\item ${\rm Im}(\Phi_\Omega(\psi))$ is the flux across the boundary $\Omega$ of destructive and constructive interferences between scattered waves.
\end{itemize}
Note that the annihilation of the complex extension of the probability flux (\ref{cprobfl}) is equivalent to the annihilation of the surface term in (\ref{surfac}).

\subsection{Quantum fluctuations inside the $[-a,a]$ interval}
We consider scalar field fluctuations distorted by $\delta-\delta^\prime$ plates and study the QFT that arises in the interval $[-a,a]$ regardless of what happens outside this interval . 
Thus $M\equiv [-a,a]\subset \mathbb{R}$, whereas its boundary is the set of the two endpoints: $\Omega=\{a,-a\}$. The complexified probability flux $\Phi_\Omega$ for a given wave function $\psi$ in this one-dimensional interval is:
\begin{equation*}
\Phi_\Omega(\psi)\equiv i\left( \psi^*(-a\uparrow)\psi^\prime(-a\uparrow)+\psi^*(a\downarrow)\psi^\prime(a\downarrow)\right) .
\end{equation*} 
For a linear combination of left-to-right-moving and right-to-left-moving plane waves with the same momentum $k$ inside $[-a,a]$ the complexified probability flux $\Phi_\Omega$ is
\begin{equation}
\Phi_\Omega=-2k\left[\left|A\right|^2-\left|B\right|^2+2i\cos(2ka){\rm Im}\left(A B^*\right)\right] \, \, .
\end{equation}
Plugging the scattering solutions of the double $\delta-\delta^\prime$ potential
\begin{equation*}
U(x)=\mu_1\delta(x+a)+\lambda_1\delta^\prime(x+a)+\mu_2\delta(x-a)+\lambda_2\delta^\prime(x-a),
\end{equation*}
we obtain the following complexified fluxes:
\begin{widetext}
\begin{eqnarray}
&&\Phi^{(R)}_\Omega=\frac{-2  k(\lambda_1^2-4)^2}{\left|\Delta\right|^2}\cdot\left[k^4(\lambda_2^2-4)^2+ 4ik^2  {\rm cos}(2ka)\left\{k(\lambda_2-2)^2\mu_2{\rm cos}(2ak)+2(k^2\lambda_2(\lambda_2^2+4)+\mu_2^2){\rm sin}(2ak)\right\}\right]\label{flux-r}\\ 
&&\Phi^{(L)}_\Omega=\frac{2  k(\lambda_2^2-4)^2}{\left|\Delta\right|^2}\cdot\left[k^4(\lambda_1^2-4)^2-4i k^2 \cos(2ka)\left\{k(\lambda_1+2)^2\mu_1\cos(2ak)+2(-k^2\lambda_1(\lambda_1^2+4)+\mu_1^2){\rm sin}(2ak)\right\}\right]\label{flux-l}
\end{eqnarray}.
\end{widetext}
In order to obtain a unitary QFT between $\delta-\delta^\prime$ plates regardless of what happens outside the space between plates we must study the conditions under which these complexified fluxes become null keeping non-trivial wave functions between plates.

\section{Robin boundary conditions versus $\delta-\delta'$ interactions}
\label{sec-robin}

The annihilation conditions $\Phi_\Omega^{(R)}(\psi)=0$, $\Phi_\Omega^{(L)}(\psi)=0$ ensure that the quantum field theory defined in the interval is unitary regardless of what happens outside this region. $\Phi_\Omega(\psi)=0$ is only compatible with a non null wave function between the plates for a discrete set of momenta $k_n$, i.e., a spectral condition giving rise to a pure point spectrum emerges when wave packets live only in the finite interval $-a<x<a$. The idea is to compare this spectrum to the spectra arising in the Asorey-Munoz-Castaneda formalism for the different boundary conditions compatible with unitary quantum field theories, see References \cite{aim,Asorey:2008xt,Munoz-Castaneda:2013yga,mc-asorey,Munoz-kk,munoz-bordag2014}, in order to identify what kind of boundary conditions can be reproduced by means of the pair of $\delta-\delta^\prime$ interactions. We remark that four real parameters define this family of potentials, whereas the self-adjoint extensions of the free particle Hamiltonian on a finite interval have deficit indices $(2,2)$ and form thus also a four-parametric family. 

Formulas (\ref{flux-r}) and (\ref{flux-l}) show that the real part of the complex flux (the probability flux across the $\delta-\delta^\prime$ plates) is given by
\begin{eqnarray}
{\rm Re}(\Phi^{(R)}_\Omega)&=&\frac{-2  k^5}{\left|\Delta\right|^2}\cdot (\lambda_1^2-4)^2(\lambda_2^2-4)^2,\label{pflux-r}\\
{\rm Re}(\Phi^{(L)}_\Omega)&=&\frac{2  k^5}{\left|\Delta\right|^2}\cdot (\lambda_1^2-4)^2(\lambda_2^2-4)^2\label{pflux-l}.
\end{eqnarray}
It is of note that whenever one of the $\delta^\prime$ couplings takes one of the values $\pm 2$ the probability fluxes 
across the interval endpoints are identically zero for both incoming from the left and from the right scattering waves
.

\subsection{Dirichlet boundary conditions}
The easiest interpretation of boundary conditions as $\delta-\delta^\prime$ interactions starts by considering the opaque limit where the two
couplings of the $\delta$ interactions $\mu_1$, $\mu_2$ are sent to $+\infty$. The analysis follows Reference \cite{Munoz-Castaneda:2013yga}
where the same limit without extra $\delta^\prime$s was analysed. Before of taking the limit we write the scattering data for incoming from the left waves and equal $\delta$-couplings $\mu_1=\mu_2=\mu$ :
\begin{eqnarray*}
 && \Delta(k;\lambda_1,\mu,\lambda_2,\mu,a)=\\ && 4 e^{4ika}\left(2k\lambda_2-i\mu\right)\left(2k\lambda_2-i\mu\right)\cdot\left(2k\lambda_1+i\mu\right)+ \\&&
 + \left(k(\lambda_1^2+4)+2i\mu\right)\cdot\left(k(\lambda_2^2+4)+2i\mu\right) =\\
 && =
 \Delta_2(k;\lambda_1,\lambda_2,a)\mu^2+ \Delta_1(k;\lambda_1,\lambda_2,a)\mu + \Delta_0(k;\lambda_1,\lambda_2,a)
\end{eqnarray*}
\begin{widetext}
\begin{eqnarray*}
&& r_R(k) = -\frac{2 e^{2 i a k}}{\Delta(k;\mu,\lambda_1,\mu,\lambda_2,a)}\cdot \left\{ [ k(\lambda_1^2+4)-2i\mu][2k\lambda_2+i\mu]+e^{-2ika}[ k(\lambda_2^2+4)+2i\mu][2k\lambda_1+i\mu]\right\} \\ && 
t(k)=\frac{(\lambda_1^2-4)(\lambda_2^2-4) k^2}{\Delta(k;\mu,\lambda_1,\mu,\lambda_2,a)} \, \, , \, \,  A_R(k)= \frac{k(\lambda_1^2-4)(k(\lambda_2^2+4)+2i\mu )}{\Delta(k;\mu,\lambda_1,\mu,\lambda_2,a)} \, \, , \, \, B_R(k)= -2\frac{k  e^{2 i a k}(\lambda_1^2-4)(2k\lambda_2+i\mu)}{\Delta(k;\mu,\lambda_1,\mu,\lambda_2,a)}
\end{eqnarray*}
\end{widetext}
For arbitrary $k>0$ one finds that the wall at $x=-a$ is completely opaque:
\begin{eqnarray*}
 && \lim_{\mu\rightarrow\infty}r_R(k)=-e^{-2 i a k};\,\, \lim_{\mu\rightarrow\infty}t(k)=0 \\&& \lim_{\mu\rightarrow\infty}A_R(k)=\lim_{\mu\rightarrow\infty}B_R(k)=0
\end{eqnarray*}
There is total reflection and it is clear that a similar pattern would be found for scattering waves incoming from the right
in the wall at $x=a$. Moreover, one may also check that the flux is zero: 
\begin{equation*}
\lim_{\mu\to +\infty}\Phi_\Omega^{(R)}=\lim_{\mu\to +\infty}\Phi_\Omega^{(L)}=0
\end{equation*}
showing that this limit is compatible with unitarity for arbitrary $k$.

There are, however, characteristic values of $k$ that allow life between the walls. Annihilation of the term proportional to $\mu^2$
in the denominator is achieved by a discrete set of momenta:
\begin{equation*}
\Delta_2(k;a)=4\left(e^{2 i k a}-1\right)(e^{2 i k a}+1)=0 \, \Leftrightarrow \, k_n=\frac{\pi}{2 a}n
\end{equation*}
where $n\in{\mathbb N}^*$ is a positive natural number. For these discrete momenta the opaque limit gives rise to non trivial solutions between plates:
\begin{eqnarray*}
&& t(k_n,\mu=\infty)=0 \quad , \quad r_R(k_n,\mu=\infty)=-(-1)^n \\
&& A_R(k_n,\mu=\infty) =\frac{4-\lambda_1^2}{8+4(\lambda_1-\lambda_2)+\lambda_1^2+\lambda_2^2}\\&& A_R(k_n,\mu=\infty) = -(-1)^n B_R(k_n,\mu=\infty)
\end{eqnarray*}
The wave functions between walls for these momenta are of the form:
\begin{eqnarray}
&&\hspace{-0.2cm}\psi(x;k_n,\mu=\infty)=\nonumber\\ &&\hspace{-0.2cm}=\frac{4-\lambda_1^2}{8+4(\lambda_1-\lambda_2)+\lambda_1^2+\lambda_2^2}\Big(e^{i k_n x}-(-1)^ne^{-i k_n x}\Big) \label{dirwf}
\end{eqnarray}
that satisfies Dirichlet boundary conditions: $\psi(\pm a;k_n,\mu=\infty)=0$. We remark that the values $\lambda_1=\pm 2$ must be excluded.
Identical pattern arises for incoming from the right scattering waves exchanging $\lambda_1$ by $\lambda_2$ because the wall at $x=a$ is also opaque in this ultra-strong limit. The same restrictions on the incoming from the right scattering waves produce the collapse to the same (\ref{dirwf}) wave functions, merely replacing $\lambda_1$ by $\lambda_2$. We should stress
that $\Delta_2(k;a)$ is the spectral function for Dirichlet boundary conditions found by Asorey and Munoz-Castaneda, whereas the ${\mathbb U}(2)$ matrix characterizing the self-adjoint extension is ${\mathbb U}=-\left(\begin{array}{cc} 1 & 0 \\ 0 & 1 \end{array}\right)$.

\subsection{Mixed Dirichlet-Neumann boundary conditions}
We consider now the effect of setting, e.g., $\lambda_2=2$. To isolate the behaviour of the system for this
critical strength of the coupling we switch-off the $\delta$-interactions altogether: $\mu_1=\mu_2=0$. Under these assumptions $A_R$ and $B_R$ become
\begin{equation*}
A_R=-\frac{\lambda_1^2-4}{4+4 e^{4 i a k}\lambda_1+\lambda_1^2} \, \, , \, \, B_R=\frac{(\lambda_1^2-4)e^{2 i a k}}{4+4 e^{4 i a k}\lambda_1+\lambda_1^2},
\end{equation*}
and the wave function between walls and its derivative read
\begin{eqnarray*}
&& \psi_k^R(x; \lambda_1,a)=\frac{\lambda_1^2-4}{4+4e^{4ika}\lambda_1+\lambda_1^2}\left(e^{-ik(x-2a)}-e^{ikx}\right)\\ && (\psi^\prime)_k^R(x; \lambda_1, a)=\frac{-ik(\lambda_1^2-4)}{4+4e^{4ika}\lambda_1+\lambda_1^2}\left(e^{-ik(x-2a)}+e^{ikx}\right)
\end{eqnarray*}
Evaluated at the endpoints the wave function and its derivative are
\begin{eqnarray*}
 && \psi_k^R(a;\lambda_1,a)=0 \\ && (\psi^\prime)_k^R(a;\lambda_1,a)=-2ik\frac{e^{ika}(\lambda_1^2-4)}{4+4e^{4ika}\lambda_1+\lambda_1^2}\\ && \psi_k^R(-a;\lambda_1,a)=\frac{(\lambda_1^2-4)e^{-ika}}{4+4e^{4ika}\lambda_1+\lambda_1^2}\left(e^{4ika}-1\right) \\ &&
(\psi^\prime)_k^R(-a; \lambda_1,a)=\frac{ike^{-ika}(4-\lambda_1^2)}{4+4e^{4ika}\lambda_1+\lambda_1^2}
\left(e^{4ika}+1\right)
\end{eqnarray*}
Discrete values of the momenta such that $e^{4ik_n a}=1$, where $k_n=\frac{\pi}{2a}n$ and $n$ is an integer, give rise to wave functions complying with Dirichlet boundary conditions at both boundary points. Discrete momenta for which $e^{4ik_n a}=-1$, with now
 $k_n=\frac{\pi}{4 a}n$, produce wave functions complying with mixed Dirichlet, on $x=a$, and Neumman, on $x=-a$, boundary conditions.
A new spectral function, $h^{(2a)}_{dn}(k)\propto 2{\rm cos}(2ak)$, corresponds to this last possibility. The complex probability flux for these values of the parameters is:
\begin{equation*}
\Phi_\Omega(\psi^{R})=-\frac{4ik(\lambda_1^2-4)^2{\rm cos}(2ka){\rm sin}(2ka)}{\vert 4+4e^{4 i a k}\lambda_1+\lambda_1^2\vert^2}
\end{equation*}
and both the momenta leading to pure Dirichlet or Dirichlet-Neumann boundary conditions are bona fide unitary quantum field theories. Exchanging 
$\lambda_2=2$ by $\lambda_1=2$ offers the same pattern on the $\psi^{(L)}$ wave functions vanishing in the opposite wall. No novelties arise 
choosing $-2$ instead $2$.

\subsection{Robin boundary conditions}
 We set first $\lambda_2=-2$  as the $\delta^\prime(x-a)$ coupling and leave free the other parameters. In the Appendix it is shown that
 \begin{equation*}
 {\bf{\rm R1}}: \psi_k^{(R)}(a\downarrow)+\frac{4}{\mu_2}\psi^{\prime(R)}_k(a\downarrow)=0 \, \,
 \end{equation*}
 i.e., the $\delta-\delta^\prime$ interaction at $x=a$ is a Robin plate towards the left if $\lambda_2=-2$. 
 The amplitudes between walls for $\lambda_2=-2$ become:
\begin{eqnarray*}
&& A_R(k;\mu_1,\lambda_1,\mu_2, -2,a)=\frac{2k(\lambda_1^2-4)(4k+i\mu_2)}{\Delta(k;\mu_1,\lambda_1,\mu_2,-2,a)}\\ && B_R(k;\mu_1,\lambda_1,\mu_2, -2,a)=-\frac{2ke^{2ika}(\lambda_1^2-4)(i\mu_2-4k)}{\Delta(k;\mu_1,\lambda_1,\mu_2,-2,a)} \\ && 
\Delta(k;\mu_1,\lambda_1,\mu_2,-2,a)=4e^{4ika}(2k\lambda_1-i\mu_1)(i\mu_2-4k)+\\ && \hspace{3.4cm}+\left(k(\lambda_1^2+4)+2i\mu_1\right)\cdot
\left(8k+2i\mu_2\right)
\end{eqnarray*}
These formulas lead to the following expression for the complex probability flux across the walls from the interior due to the \lq\lq diestro\rq\rq scattering wave functions $\psi_k^{(R)}$ in $-a<x<a$:
\begin{eqnarray*}
 && \Phi_\Omega(\psi_k^{(R)};\mu_1,\lambda_1,\mu_2,-2,a)=\frac{128 i k^2(\lambda_1^2-4)^2}{\vert\Delta (k,\mu_1,\lambda_1,\mu_2,-2,a)\vert^2}\\&& \hspace{1cm}\times\, \, {\rm cos}(2ka)\left\{ k\mu_2{\rm cos}(2ka)-2(k^2-\frac{\mu_2^2}{16}){\rm sin}(2ka)\right\}  \, \, .
\end{eqnarray*}
The complex flux $\Phi^{(R)}$ is null for the values of momenta that solve the following transcendent equation:
\begin{equation}
k_n\mu_2{\rm cos}(2k_na)-2(k_n^2-\frac{\mu_2^2}{16}){\rm sin}(2k_na)=0  \, \,  \label{robbcsc}
\end{equation}

In the Asorey-Munoz-Castaneda formalism the boundary conditions compatible with an unitary quantum field theory in $(1+1)$-dimensions are characterized by an ${\mathbb U}(2)$ matrix connecting the values of the scalar field fluctuations at the two points of the boundary:
\begin{eqnarray*}
&&\left(\begin{array}{c} \phi(t_0,-a)+i\phi^\prime(t_0,-a) \\ \phi(t_0,a)-i\phi^\prime(t_0,a)\end{array}\right)=\\&&\hspace{2cm}=\mathbb{U}(\alpha,\beta,\theta,\gamma)\left(\begin{array}{c} \phi(t_0,-a)-i\phi^\prime(t_0,-a) \\ \phi(t_0,a)+i\phi^\prime(t_0,a)\end{array}\right)
\end{eqnarray*}
The matrix 
\begin{equation*}
\mathbb{U}(\alpha,\beta,\theta,\gamma)=e^{i\alpha}\left({\rm cos}(\beta).\mathbb{I}+i{\rm sin}(\beta)\vec{n}(\theta,\gamma)\vec{\sigma}\right)
\end{equation*}
depends on four angles: $\alpha\in[0,2\pi]$, $\beta\in[-\frac{\pi}{2},\frac{\pi}{2}]$, $\theta\in[0,\pi]$, $\gamma\in[0,2\pi]$, $\vec{n}(\theta,\gamma)$ is a vector varying on a $S^2$ sphere of radius $1$, and the components of $\vec{\sigma}$ are the Pauli matrices. Robin boundary conditions correspond to $\beta=0$ such that the ${\mathbb U}(2)$ matrix becomes $\mathbb{U}(\alpha)= e^{2i\alpha}\mathbb{I}$. Momenta compatible with these Robin boundary conditions belong to the kernel of the spectral function:
\begin{equation}
 h_R(k;\alpha)\propto \left\{\begin{array}{c} -2 k {\rm sin}(2\alpha){\rm cos}(2ka)+\\ +2\left(k^2a^2{\rm cos}^2(\alpha)-{\rm sin}^2(\alpha)\right){\rm sin}(2ka)\end{array}\right. \label{robbcsf} \, .
\end{equation}
The spectral conditions (\ref{robbcsc}) and (\ref{robbcsf}) are identical iff ${\rm tan}\alpha =\frac{\mu_2 a}{4}$ and Robin boundary conditions 
are obtained in the two $\delta/\delta^\prime$ plates.

In fact, if $\lambda_2=-2$ the wave functions between walls and their derivative read:
\begin{eqnarray*}
\psi_k^{(R)}(x)&=&\frac{-2 k(\lambda_1^2-4)\left[e^{-ik(x-2a)}\frac{i\mu_2-4k}{i\mu_2+4k}-e^{ikx}\right]}{\Delta(k;\mu_1,\lambda_1,\mu_2,-2,a)}\\
\psi^{\prime(R)}_k(x)&=&\frac{2ik^2 (\lambda_1^2-4)\left[e^{-ik(x-2a)}\frac{i\mu_2-4k}{i\mu_2+4k}+e^{ikx}\right]}{\Delta(k;\mu_1,\lambda_1,\mu_2,-2,a)}
\end{eqnarray*}
such that their values at the boundary points are:
\begin{eqnarray*}
&&\psi_k^{(R)}(a\downarrow)=2\frac{k^2e^{ika}(\lambda_1^2-4)}{\Delta(k;\mu_1,\lambda_1,\mu_2,-2,a)} \\ && \psi^{\prime(R)}_k(a\downarrow)=-4\frac{k^2e^{ika}(\lambda_1^2-4)\mu_2}{\Delta(k;\mu_1,\lambda_1,\mu_2,-2,a)}\\ && \psi_k^{(R)}(-a\uparrow)=\frac{-\mu_2 {\rm cos}(2ak)+4k{\rm sin}(2ak)}{8}\psi^{(R)}_k(a\downarrow) \\ && \psi^{\prime(R)}_k(-a\uparrow)=-\frac{4k {\rm cos}(2ak)+\mu_2{\rm sin}(2ak)}{\mu_2}\psi_k^{\prime(R)}(a\downarrow)
\end{eqnarray*}
The linear combinations of the wave function and its derivative at the $x=\pm a$ points satisfy:
\begin{eqnarray*}
{\bf{\rm R1}}: && \, \, \, \psi_k^{(R)}(a\downarrow)+\frac{4}{\mu_2}\psi^{\prime(R)}_k(a\downarrow)=0  \\ {\bf{\rm R2}}: && \, \, \, \psi_k^{(R)}(-a\uparrow)+\\ && \, \, \, +\frac{4 k{\rm cos}(2ka)+\mu_2{\rm sin}(2ka)}{k(\mu_2{\rm cos}(2ka)-4 k{\rm sin}(2ka))}\psi^{\prime(R)}_k(-a\uparrow)=0 \, \, . 
\end{eqnarray*}
Plugging in ${\bf{\rm R2}}$ the momenta complying with the spectral condition (\ref{robbcsc}) one finds
\begin{equation*}
\frac{4 k_n{\rm cos}(2k_na)+\mu_2{\rm sin}(2k_n a)}{k_n(\mu_2{\rm cos}(2k_na)-4 k_n{\rm sin}(2k_na))}=-\frac{4}{\mu_2} 
\end{equation*}
such that the plate at $x=-a$ is also a Robin plate for these discrete set of wave numbers. In sum, the quantum fluctuations corresponding to 
\lq\lq diestro\rq\rq scattering satisfy Robin boundary conditions if $\lambda_2=-2$ and the momenta are selected by the spectral condition
(\ref{robbcsc}).

To build Robin boundary conditions also on the \lq\lq zurdo\rq\rq scattering wave functions $\psi_k^{(L)}$ we follow an identical process setting $\lambda_1=2$ as the $\delta^\prime(x+a)$ coupling and leaving free all the other three. In the Appendix we have shown that:
\begin{equation*}
 {\bf{\rm R2}}: \psi_k^{(L)}(-a\uparrow)-\frac{4}{\mu_1}\psi^{\prime(L)}_k(-a\uparrow)=0 \, \,
 \end{equation*}
The annihilation of the complex flux towards the exterior of the \lq\lq zurdo\rq\rq scattering waves between plates
\begin{eqnarray*}
&& \Phi_\Omega(\psi_k^{L};\mu_1,2,\mu_2,\lambda_2,a)=\frac{128 i k^2(\lambda_2^2-4)^2}{\vert\Delta (k,\mu_1,2,\mu_2,\lambda_2,a)\vert^2}\\&& \hspace{1cm}\times\, \, {\rm cos}(2ka)\left\{ k\mu_1{\rm cos}(2ka)-2(k^2-\frac{\mu_1^2}{16}){\rm sin}(2ka)\right\} \, .
\end{eqnarray*}
is accomplished by quantum fluctuations with wave numbers satisfying the spectral condition
\begin{equation}
k_n\mu_1{\rm cos}(2k_na)-2(k_n^2-\frac{\mu_1^2}{16}){\rm sin}(2k_na)=0  \, \,  \label{robbcscl} \, \, .
\end{equation}
For this discrete set of momenta complying with (\ref{robbcscl}) $A_L$ and $B_L$ are non null and fluctuations that do not cross the walls survive in the interior. Repeating the arguments above we find that the $\psi_k^{(L)}$ waves satisfy also Robin boundary conditions at $x=a$:
\begin{eqnarray*}
{\bf{\rm R1}}: && \, \, \, \psi_k^{(L)}(a\downarrow)+\frac{4}{\mu_1}\psi^{\prime(L)}_k(a\downarrow)=0  \\ {\bf{\rm R2}}: && \, \, \, \psi_k^{(L)}(-a\uparrow) -\frac{4}{\mu_1}\psi^{\prime(L)}_k(-a\uparrow)=0 \, \, . 
\end{eqnarray*}

One might wonder about the simultaneous choice of $\lambda_1=2$ and $\lambda_2=-2$ as $\delta^\prime(x+a)$ and $\delta^\prime(x-a)$ couplings. 
A new set of Robin boundary conditions arise at this plane in the parameter space: 
\begin{eqnarray}
{\bf{\rm R1}}: && \, \, \, \psi_k(a\downarrow)+\frac{4}{\mu_2}\psi_k^\prime(a\downarrow)=0  \nonumber\\ {\bf{\rm R2}}: && \, \, \, \psi_k(-a\uparrow) -\frac{4}{\mu_1}\psi^{\prime}_k(-a\uparrow)=0 \, \, . \label{robrob}
\end{eqnarray}
Plane waves of the form $\psi_k(x)=A e^{i k x}+ B e^{-i k x}$, solutions of our Hamiltonian inside $[-a,a]$, satisfy the boundary conditions (\ref{robrob}) if the determinant of the corresponding homogeneous linear system
\begin{eqnarray}
&& {\rm det}\left(\begin{array}{cc} e^{ i k a}(1+4i\frac{k}{\mu_2}) & e^{- i k a}(1-4i\frac{k}{\mu_2}) \\ e^{- i k a}(1-4i\frac{k}{\mu_1}) & e^{i k a}(1+4i\frac{k}{\mu_1})\end{array}\right)=\label{detrob}\\ && = -\frac{16i}{\mu_1\mu_2}\left[2(k^2-\frac{\mu_1\mu_2}{16})\sin (2 a k)-k\frac{\mu_1+\mu_2}{2}\cos (2 a k)\right]\nonumber
\end{eqnarray}
is zero. The solution is:
\begin{eqnarray}
&& A(k_n)=1 \, \, , \, \, B(k_n)=e^{2 i k_n a}\frac{\mu_2+4 i k_n}{\mu_2´-4 i k_n} \nonumber \\ && \tan(2 a k_n)=\frac{4 k_n(\mu_1+\mu_2)}{16 k_n^2-\mu_1\mu_2} \label{robwf} \quad .
\end{eqnarray}

From the point of view of scattering waves both walls respectively at $\lambda_1=2$  and $\lambda_2=-2$ are opaque to waves of generic wave number $k$ and there is nothing between plates: $A_R(k,\mu_1,2,\mu_2,-2,a)= A_L(k,\mu_1,2,\mu_2,-2,a)=B_R(k,\mu_1,2,\mu_2,-2,a)=B_L(k,\mu_1,2,\mu_2,-2,a)=0$. A discrete subset of momenta, however, escapes this fate. The discriminant is in this case
\begin{eqnarray}
&& \Delta(k,\mu_1,2,\mu_,-2,a)=\label{discr}\\ && = 64 i e^{2 ik a}\left[k \frac{\mu_1+\mu_2}{2} \cos(2 k a)-2(k^2-\frac{\mu_1 \mu_2}{16})\sin(2 k a)\right]\nonumber
\end{eqnarray}
such that non-zero wave functions inside the $[-a,a]$ interval are still possible for all those wave numbers $k_n$
belonging to the kernel of $\Delta$. A third spectral condition (\ref{robwf}) arises annihilating both (\ref{detrob}) and (\ref{discr}). Moreover, a collapse of the \lq\lq diestro\rq\rq and \lq\lq zurdo\rq\rq states to a single wave function satisfying Robin boundary conditions is caused by the choice $\mu_1=\mu_2=\mu$ to find the same situation as described above for Dirichlet boundary conditions.

\section{Quantum vacuum interaction between two $\delta-\delta'$ plates}
\label{sec:6}
In this Section we consider the two $\delta-\delta^\prime$ potentials as modelling two monoatomic infinitely thin plates with its physical properties encoded in the $\mu$ and $\lambda$ couplings. The goal is the computation of the quantum vacuum interaction energies between these partially transparent plates.
\subsection{$T$-operator for parity breaking point potentials}

The reduced Green's function (\ref{red-green-scatt}) for a potential concentrated at $x=0$ is of the general form:
\begin{widetext}
\begin{equation}
  G_\omega(x,y)=G_\omega^{(0)}(x-y)+\left\{
                             \begin{array}{cc}
                             \delta G_\omega^{(-,-)}(x,y)=-\frac{r_R(k)}{2ik}e^{-ik(x+y)} & \hspace{-1cm}, \, x,y<0\\
                             \delta G_\omega^{(+,+)}(x,y)=-\frac{r_L(k)}{2ik}e^{ik(x+y)} & \hspace{-1cm}, \, x,y>0\\
                            \delta G_\omega^{(\mp,\pm)}(x,y)=-\frac{(t(k) -1)}{2ik}e^{ik|x-y|} & , \, {\rm sgn}(x y)=-1 
                            \end{array}\right. \, \, , \, \,
                          G_\omega^{(0)}(x-y)=-\frac{1}{2ik}e^{ik|x-y|}  \label{point-prop}\, \, \,  .
\end{equation}
\end{widetext}
The formula obtained in sub-section IV.A of Reference \cite{Munoz-Castaneda:2013yga} for the integral kernel of the $T$-operator 
\begin{equation}
T_\omega(x,y)=2ik(t(k)-1)\delta(x)\delta(y),\label{t-op-pointeven}
\end{equation}
in terms of the transmission amplitude $t(k)$ works fine for parity even point interactions producing unimodal scattering processes. In such a case $r_R(k)=r_L(k)=t(k)-1$, see e.g. Ref. \cite{galindo1990quantum}. In these cases the reduced Green function (\ref{point-prop}) is such that
\begin{equation*}
-\frac{r(k)}{2ik}=\lim_{x\uparrow 0,y\uparrow 0}\delta G_\omega^{(-,-)}(x,y)=\lim_{x\downarrow 0,y\downarrow 0}\delta G_\omega^{(+,+)}(x,y) \, .
\end{equation*}
$\delta G_\omega^{(\mp,\pm)}(x,y)$ are not differentiable at the origin and produce through their second derivatives at $x=y=0$ the even point potential. The delta function source in the equation for the Green function is accounted for by the second derivative of $G^{(0)}_\omega(x-y)$ at $x=y=0$. 

When the point potential is not parity even, $r_R(k)\neq r_L(k)$ and $t(k)-1\neq r_{R,L}(k)$. We have
\begin{eqnarray*}
-\frac{r_R(k)}{2ik}&=&\lim_{x\uparrow 0,y\uparrow 0}\delta G_\omega^{(-,-)}(x,y) \\ -\frac{r_L(k)}{2ik}&=&\lim_{x\downarrow 0,y\downarrow 0}\delta G_\omega^{(+,+)}(x,y) \, .
\end{eqnarray*}
There is a step discontinuity at the origin between $\delta G_\omega^{(-,-)}(x,y)$ and $\delta G_\omega^{(+,+)}(x,y)$ responsible, through the second derivative, of a parity odd point interaction. The parity even point potential is due to the same part of the Green function as before.
Together with the delta source coming from $G_\omega^{(0)}$ these Dirac delta and delta' will saturate the Green equation. We conclude that  $\delta G_\omega^{(-,-)}(x,y)$ and $\delta G_\omega^{(+,+)}(x,y)$ contribute to the $T$-operator of a point interaction neither even nor odd.  For point potentials that do not have well-defined parity one considers the following directional limits:
\begin{eqnarray*}
t(k)-1&=&-2ik\lim_{x\downarrow 0,y\uparrow 0}\delta G_\omega^{(+,-)}(x,y) \\
t(k)-1 &=&-2ik\lim_{x\uparrow 0,y\downarrow 0}\delta G_\omega^{(-,+)}(x,y)\\ r_R(k)&=&-2ik\lim_{x\uparrow 0,y\uparrow 0}\delta G_\omega^{(-,-)}(x,y)\\
r_L(k)&=&-2ik\lim_{x\downarrow 0,y\downarrow 0}\delta G_\omega^{(+,+)}(x,y) \, \, \, .
\end{eqnarray*} 
Collectively denoting  the difference between the Green functions of ${\bf K}$ and ${\bf K}^{(0)}$ in the four quadrants of the real line times itself as $\delta G_\omega(x,y)$ and using $f(x)\delta(x)=f(0)\delta(x)$ one immediately generalizes equation (\ref{t-op-pointeven}) to obtain the $T$-operator for a point interaction neither even nor odd:
\begin{equation}
T_\omega(x,y)=-(2ik)^2\delta G_\omega(x,y)\delta(x)\delta(y),\label{t-op-pointgen}
\end{equation}
Accordingly, the $T$-operator expressions in terms of the scattering amplitudes are different in distinct quadrants of the $(x,y)$-plane: 
\begin{equation}
T_\omega(x,y)=2ik\delta(x)\delta(y)\left\{
                             \begin{array}{cc}
                             r_R &, x,y<0\\
                             r_L &, x,y>0\\
                            t -1 &, {\rm sgn}(x y)=-1
                             \end{array}\right. .\label{t-op-pointgen2}
\end{equation}
In the first and third quadrant the $T$-matrix formula picks the reflection amplitudes, whereas in the second and four quadrant the transmission
amplitude is pertinent. It is clear that for even unimodal point potentials all the components are equal. We write the $T$-operator in the compact form
\begin{eqnarray*}
&& T_\omega(x,y)=\\&=&\theta(x)\theta(y)T^{(+,+)}_\omega(x,y)+\theta(-x)\theta(-y)T_\omega^{(-,-)}(x,y)\nonumber \\
&+&\theta(x)\theta(-y)T_\omega^{(+,-)}(x,y)+\theta(-x)\theta(y)T_\omega^{(-,+)}(x,y)
\end{eqnarray*}
such that the different components of the $T$-operator are identified from equation (\ref{t-op-pointgen2}). 

\subsection{The $TGTG$-formula for parity breaking point potentials}

In Reference \cite{Kenneth:2007jk}, subsection VII.A , it is shown that only the $T^{(+,+)}_\omega$ and $T^{(-,-)}_\omega$ components of the $T$-operator enter in $TGTG$ formula of the quantum vacuum interaction energy between two compact objects. The basis of the $TGTG$-formula is the $M$ operator:
\begin{equation*}
 {\bf M}_\omega= {\bf G}^{(0)}_\omega {\bf T}_{\omega(1)}^{(+,+)}{\bf G}^{(0)}_\omega {\bf T}_{\omega(2)}^{(-,-)} \, .
\end{equation*}
In this context we have:
\begin{eqnarray}
T_{\xi(1)}^{(+,+)}(z_1,z_2)&=&-2\kappa \, r_L^{(1)}(\kappa)\delta(z_1+a)\delta(z_2+a)\label{tma1}\\
T_{\xi(2)}^{(-,-)}(z_1,z_2)&=&-2\kappa \, r_R^{(2)}(\kappa)\delta(z_1-a)\delta(z_2-a) \, , \label{tma2}
\end{eqnarray}
where we have performed the Wick rotation $\omega\to i\xi$ and considered the energy-momentum dispersion relation of a massless scalar particle: $\omega=k \to
i\xi\,=i\kappa$. The $T$ integral kernel in formula (\ref{tma1}) is due to the $(\mu_1,\lambda_1)$ point interaction located at $x=-a$ relating points such that $z_1,z_2>-a$. The $T$ integral kernel in (\ref{tma2}) comes from the $(\mu_2,\lambda_2)$ point interaction located at $x=a$ relating points such that $z_1,z_2<a$.

We write now the $TGTG$-formula in terms of the associated integral kernels after the Wick rotation in the form:
\begin{eqnarray*}
&& M_\xi(x,y)= \int_{-a}^\infty\, dz_1\int_{-a}^\infty\, dz_2\int_{-\infty}^a\, dz_3 \,\cdot \\ &&\cdot\left[ G_\xi^{(0)}(x,z_1)T^{(+,+)}_{\xi(1)}(z_1,z_2) 
\cdot G_\xi^{(0)}(z_2,z_3)T^{(-,-)}_{\xi(2)}(z_3,y)\right]\\ && =r_L^{(1)}(\kappa)r_R^{(2)}(\kappa)e^{-2\kappa a}e^{-\kappa\vert x+a\vert}\delta(y-a)\, \, \, .
\end{eqnarray*}
From this kernel we obtain the trace of the ${\bf M}_\xi$-operator taking $x=y$ and integrating over the interval $[-a,a]$ where these two points may coincide:
\begin{equation*}
{\rm Tr}_{L^2}{\bf M}_\xi=\int_{-a}^a\, dx \, M_\xi(x,x)=r_L^{(1)}(\kappa)r_R^{(2)}(\kappa)e^{-4\kappa a} \, .
\end{equation*}
Other combinations of the $T$-operator components do not allow full coincidences of the $x$ and $y$ points in the interval $[-a,a]$; this is the only contribution to the Euclidean  $TGTG$-formula for the quantum vacuum interaction induced by the two point potentials. The formal series expansion
\begin{eqnarray*}
&& {\rm Tr}_{L^2}\ln\left(1-{\bf M}_\xi\right)=\\
&&=-\sum_{n=1}^\infty\,\frac{(-1)^n}{n}\left[{\rm Tr}_{L^2}{\bf M}_\xi\right]^n=\ln\left(1-{\rm Tr}_{L^2}{\bf M}_\xi\right)
\end{eqnarray*}
leads to the final formula:
\begin{eqnarray}
  E_0^{{\rm int}}&=&\frac{1}{2}\int_0^\infty\frac{d\xi}{\pi}{\rm Tr}_{L^2}\ln\left(1-{\bf M}_{\xi}\right)\nonumber\\
&=&\frac{1}{2}\int_0^\infty\frac{d\xi}{\pi}\ln\left(1-r^{(1)}_L(\xi)r^{(2)}_R(\xi)e^{-4\xi a}\right)\label{tgtgeopp}\, \, .
\end{eqnarray}
Application to the ${\bf U}=\lambda\delta^\prime(x)+\mu\delta(x)$ potential provides the quantum vacuum interaction between two $\delta-\delta^\prime$ plates in terms of the right and left reflection amplitudes previously obtained:
 \begin{eqnarray}
&& E_0^{{\rm int}}=\int_0^\infty\frac{d\xi}{2\pi}\times\label{tgtgqveddp}\\ &&\ln \left(1+\frac{4 e^{-4 a \xi } \left(2 \lambda _1 \xi - \mu _1\right) \left(2 \lambda _2 \xi +\mu _2\right)}{\left[\left(\lambda _1^2+4\right) \xi +2 \mu
   _1\right] \left[\left(\lambda _2^2+4\right) \xi +2 \mu _2\right]}\right)\nonumber
\end{eqnarray}
It is of note that the quantum vacuum interaction energy depends on the non-dimensional parameters $\mu_1 a$, $\mu_2 a$, $\lambda_1$, $\lambda_2$ and is proportional to $a^{-1}$.Therefore without loss of generality we will only perform the numerical calculations setting $a=1$ which is enough to obtain information about the dependence of the vacuum energy in the $\delta-\delta^\prime$ couplings.

\subsection{Numerical and analytical evaluations}
The integral in formula (\ref{tgtgqveddp}) is not amenable to analytic integration for generic values of $\mu$ and $\lambda$. Numerical integration, however, of the right member of (\ref{tgtgqveddp}) is stable and very robust due to the decreasing exponential factor. We find that quantum vacuum energies between two $\delta-\delta^\prime$ plates can be positive, negative or zero when the parameters vary: there are repulsive, attractive, and null Casimir forces. In Figure \ref{fig3}(left) we plot the numerical results for the vacuum quantum energy in the 2D sub-space of the parameter space where the couplings at $x=-a$ and $x=a$ are identical. The quantum vacuum energy in the case of  $\lambda_1=\lambda_2=\lambda, \mu_1=\mu_2=\mu$ is numerically estimated and the pattern is shown in Figure \ref{fig3} (left) where the zones of positive, respectively negative, Casimir energy are shown by means of colors towards the infrared, respectively the ultraviolet, ends of the visible light spectrum separated by a black curve of zero energy. This pattern of the Casimir energy seems to be in contradiction with the Kenneth-Klich theorem: \lq\lq opposites attract\rq\rq, see ref. \cite{kenneth06}. The apparent contradiction is explained, however, by noticing that the double $\delta-\delta^\prime$ system of plates is not symmetric under reflection even when both plates have identical parameters. The exchange symmetry occurs only for the pure double $\delta$ system. We showed that two identical $\delta$-plates do indeed attract each other in \cite{Munoz-Castaneda:2013yga}. A natural loophole is a refinement of the inteligence of \lq\lq opposite\rq\rq: 
two $\delta-\delta^\prime$  plates are opposite if they have identical $\delta$-couplings, $\mu_1=\mu_2$, but $\delta^\prime$ couplings of opposite sign and the same modulus: $\lambda_1=-\lambda_2$. With this proviso our results fit well into the Keneth-Klicht paradigm.

\begin{widetext}
\begin{center}
\begin{figure}[h]
\center{\includegraphics[width=8.55cm]{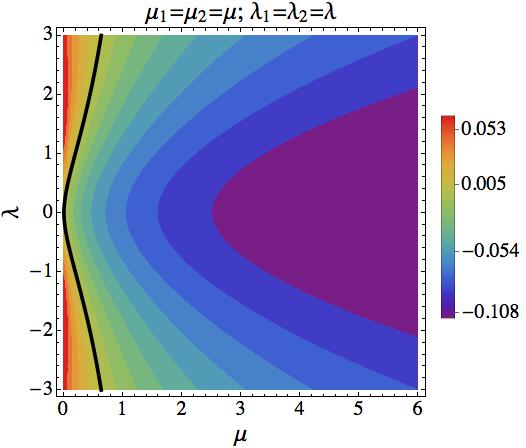}\includegraphics[width=8.25cm]{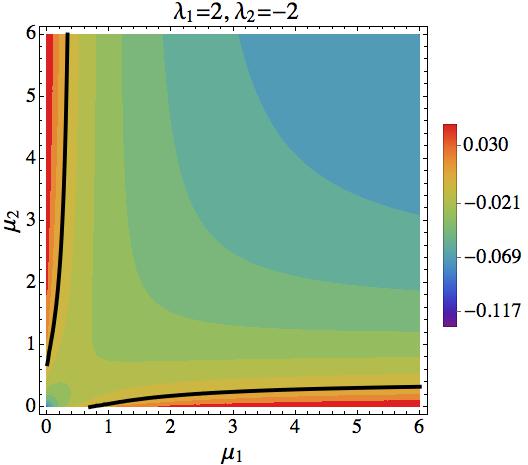}}
\caption{\footnotesize{Quantum vacuum interaction energy for $a=1$: two identical $\delta-\delta^\prime$ plates (left). $\lambda_1=-\lambda_2=2$ opaque points of different sign prompting Robin boundary conditions (right). In both cases the thick black line is the zero energy line that separates the zones of attractive and repulsive Casimir forces. Note that opposite plates that appear in the plot in the right on the diagonal $\mu_1=\mu_2$ always attract.}}
\label{fig3}
\end{figure}
\end{center}
\begin{center}
\begin{figure}[htpb]
\center{\includegraphics[width=8.55cm]{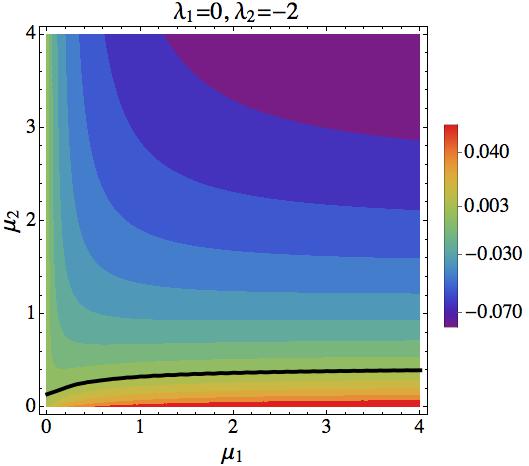}\includegraphics[width=8.69cm]{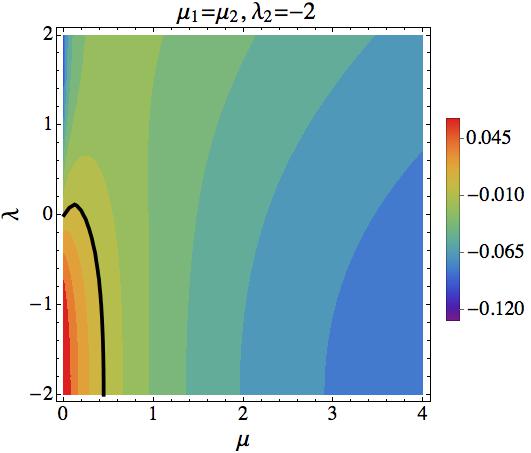}}
\caption{\footnotesize{Quantum vacuum energies for $a=1$ and $\lambda_2=-2$: the other $\delta^\prime$ turned off (left) two equal $\delta$s that allow Robin boundary conditions on the straight line $\lambda=2$, giving rise to negative vacuum energy. (right). The thick black lines denote the zero energy curves}}
\label{fig6}
\end{figure}
\end{center}
\end{widetext}

{\bf Robin quantum vacuum energy.}  
A similar structure is shown in Figure \ref{fig3} (right) by the numerically evaluated quantum vacuum energy in the two-dimensional sub-space $\lambda_1=2$, $\lambda_2=-2$. Positive, negative, and zero quantum vacuum energies also arise prompting repulsive, attractive and null forces
in this subspace corresponding to Robin boundary conditions. In Figure \ref{fig6} we plot the map of the quantum vacuum energies at the opaque point $\lambda_2=-2$ in other two cases: $\lambda_1=0$ (left) and $\mu_1=\mu_2$ (right). The quantum vacuum interaction energy follows the same pattern as in the preceeding regimes: Casimir forces appear attractive, repulsive, and null.

{\bf Two analytic evaluations: the limits $\mu_1=\mu_2=+\infty$ and $\mu_1=\mu_2=0$.} 
In the case that the two $\delta$s are infinitely repulsive $\mu_1=\mu_2=+\infty$ the integral in formula (\ref{tgtgqveddp}) is analytically
computable and we find:
\begin{equation*}
E_0^{{\rm int}}=\int_0^\infty\frac{d\xi}{2\pi}\cdot\ln \left(1- e^{-4 a \xi} \right)=-\frac{1}{12}\cdot\frac{\pi}{4a}
\end{equation*}
the well known result for Dirichlet boundary conditions on an interval of length $2a$.

In the second limit when the Dirac-$\delta$ point potentials are switched off, $\mu_1=\mu_2=0$ the integration in the $TGTG$ formula (\ref{tgtgqveddp}) can be computed analytically to find
\begin{equation}
E_0^{{\rm int}}(\lambda_1,\lambda_2)=-\frac{\text{Li}_2\left(r_L\left(0,\lambda _1\right) r_R\left(0,\lambda _2\right)\right)}{8 \pi a}
\end{equation}
\begin{center}
\begin{figure}[htbp]
\center{\includegraphics[width=8.25cm]{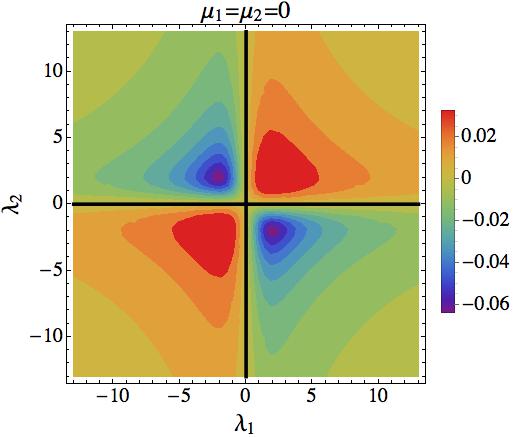}}
\caption{\footnotesize{Two-$\delta^\prime$ quantum vacuum interaction energy for $a=1$. The thick black line corresponds to zero quantum vacuum energy.}}
\label{fig5}
\end{figure}
\end{center}
where $\text{Li}_2(z)=\sum_{k=1}^\infty z^k/k^2$ is the polylogarithm of order 2 and $r_L(0,\lambda_1)$, $r_R(0,\lambda_2)$ are the left and right reflection amplitudes for a single $\delta^\prime$ respectively at $x=-a$ and $x=a$. 
The sign of the quantum vacuum interaction energy in this case is just ${\rm sgn}(\lambda_1 \lambda_2)$ and the zero energy lines are the abscissa and ordinate axes, see Figure \ref{fig5}. This result is remarkable: one needs to choose $\delta^\prime$ couplings of opposite sign to compensate the parity breaking and find attractive Casimir forces, confirming the need of refinement of the concept of opposite objects as explained above.

\subsection{The ${\rm TGTG}$ formula in higher dimensions}
We pass to study the quantum vacuum interaction energy between two $ D-1$-dimensional hyper-cubical plates. The $D-1$ plates are very thin in their perpendicular direction. If $\hat{\bf n}$ is a normal vector orthogonal to the plate we characterize them by the potential $V({\bf x}\cdot\hat{\bf n})=\mu\delta({\bf x}\cdot\hat{\bf n})+\lambda \delta^\prime({\bf x}\cdot\hat{\bf n})$ (see \footnote{Note that to obtain the $D$-dimensional version of the double $\delta-\delta^\prime$ potential the election of the vector $\hat{\bf n}$ must be the same for both $\delta-\delta^\prime$ potentials. Therefore if $\hat{\bf n}$ is any modulus one vector orthogonal to both plates the double $\delta-\delta^\prime$ potential in $D$-dimensions would be written as $V({\bf x}\cdot\hat{\bf n})=\mu_1\delta(({\bf x}+{\bf a})\cdot\hat{\bf n})+\lambda_1 \delta^\prime(({\bf x}+{\bf a})\cdot\hat{\bf n})+\mu_2\delta(({\bf x}-{\bf a})\cdot\hat{\bf n})+\lambda_2 \delta^\prime(({\bf x}-{\bf a})\cdot\hat{\bf n})$ where the vector ${\bf a}$ is parallel to $\hat{\bf n}$.}). In the ${\rm TGTG}$-formula for the quantum vacuum energy per unit \lq\lq surface\rq\rq $E_0^{{\rm int}}/\Sigma_D$, $\Sigma_D$ denoting the volume of the plate, $\kappa=\xi$ is replaced by the modulus of the momentum vector $i k=i\vert {\bf k}\vert$, such that integration becomes
$D$-dimensional and a factor containing the Euler Gamma function $\Gamma(z)$ appears:
\begin{equation*}
\int_0^\infty\frac{d\xi}{\pi}\mapsto\int_{\mathbb{R}^{D}}\frac{d^{D}{\bf k}}{(2\pi)^{D}}=\frac{ \pi ^{-D/2}}{2^{D}\Gamma \left(\frac{D}{2}\right)}\int_0^\infty dk k^{D-1} \, \, \, .
\end{equation*}
The integrand in the ${\rm TGTG}$ formula only depends on the modulus $k$, and the vacuum energy density reads:
\begin{eqnarray}
  \frac{E_0^{{\rm int}}}{\Sigma_D}&=&\frac{\pi ^{-D/2}}{2^{D} \Gamma \left(\frac{D}{2}\right)}\cdot \nonumber\\
  &&\hspace{-1cm}\cdot \int_0^\infty dk k^{D-1}\ln\left(1-r^{(1)}_L\left(ik\right)r^{(2)}_R\left(ik\right)e^{-4 ak}\right).\label{qve-d} 
\end{eqnarray}
Obviously, the integral here is much more complicated but numerical integration will remain stable due to the presence of the decreasing exponential factor. We anticipate now that the qualitative effect of higher dimensional plates on scalar field fluctuations follows a similar 
pattern to the pattern previously described for the one-dimensional set up. Setting the values of the $\delta^\prime$-couplings to be  $\lambda_1=-2=-\lambda_2$ the argument of the logarithm in equation (\ref{qve-d}) becomes:
\begin{equation*}
1-\left.r^{(1)}_L\left(ik\right)\right\vert_{\lambda_1=-2}\left.r^{(2)}_R\left(ik\right)\right\vert_{\lambda_2=2}e^{-4 ak}=1-e^{-4 a k}\, ,
\end{equation*}
the well known integrand produced by Dirichlet boundary conditions on scalar quantum fluctuations. In this case the integral in formula (\ref{qve-d}) can be computed exactly to find, for instannce for $D=3$,
\begin{equation*}
E_0^{{\rm int}}(\lambda_1=-2,\lambda_2=2)/\Sigma_D=-\frac{\pi^2}{1440}\cdot\frac{1}{ (2a)^3} \, ,
\end{equation*} 
which is a very well known result for the quantum vacuum energy interaction between $3D$ Dirichlet plates, see for example the References \cite{mc-asorey,rom-jpa35}. This calculation confirms what we show in the appendix \ref{append}, the $-\lambda_1=2=\lambda_2$ combination of $\delta^\prime$
couplings is tantamount to Dirichlet boundary conditions. We also show that the alternative combination $\lambda_1=2$ and $\lambda_2=-2$ compells the quantum vacuum fluctuations between plates to behave as if the $\delta-\delta^\prime$ plates where Robin plates, see Reference \cite{rom-jpa35} to find a direct description of this type of boundary conditions on scalar fields. For this choice $\lambda_1=2=-\lambda_2$ the argument of the logarithm in formula (\ref{qve-d}) 
\begin{eqnarray*}
&&1-\left.r^{(1)}_L\left(ik\right)\right\vert_{\lambda_1=2}\left.r^{(2)}_R\left(ik\right)\right\vert_{\lambda_2=-2}e^{-4 ak}=\\
&&1-\frac{e^{-4 a k} (4 k-\text{$\mu _1$}) (4 k-\text{$\mu_2 $})}{(4 k+\text{$\mu_1 $}) (4 k+\text{$\mu _2$})}
\end{eqnarray*}
leads to the following quantum vacuum energy per unit surface integral formula for $D$-dimensional $\delta-\delta^\prime$-plates at the Robin combination $\lambda_1=2=-\lambda_2$ of opaque points:
\begin{eqnarray}
  \frac{E_0^{{\rm int}}}{\Sigma_D}&=&\frac{\pi ^{-D/2}}{2^{D} \Gamma \left(\frac{D}{2}\right)}\cdot\nonumber\\
  &&\hspace{-1.5cm}\cdot\int_0^\infty dk k^{D-1}\ln\left(1-\frac{e^{-4 a k} (4 k-\text{$\mu_1 $}) (4 k-\text{$\mu_2 $})}{(4 k+\text{$\mu_1 $}) (4 k+\text{$\mu_2 $})}\right).\label{robqvepus}
\end{eqnarray}
Partial integration in the integral formula (5.20) of Reference \cite{rom-jpa35} tells us that this formula is identical to (\ref{robqvepus}) provided that the $b_i$, $i=1,2$, non dimensional parameters entering in formula (5.20) of Ref. \cite{rom-jpa35} are traded by: $-\frac{4}{\mu_i}\cdot\frac{1}{2a}$. We now offer two sets of graphics, Figures 6 and 7, where the quantum vacuum energy densities obtained by numerical integrations of formula (\ref{qve-d}) are plotted both for \lq\lq opposite\rq\rq and \lq\lq identical\rq\rq $\delta-\delta^\prime$ plates in 
$D=3$.

\begin{widetext}
\begin{center}
\begin{figure}[h]
\center{\includegraphics[width=8.55cm]{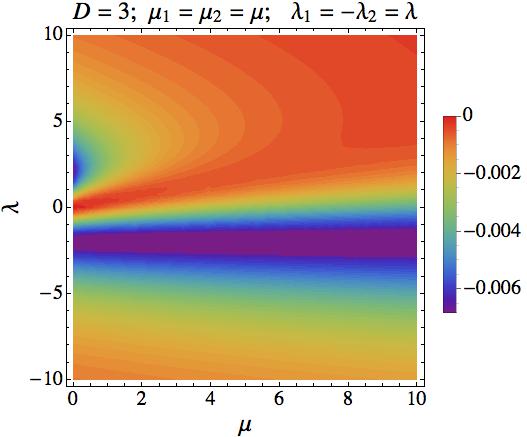}\includegraphics[width=8.25cm]{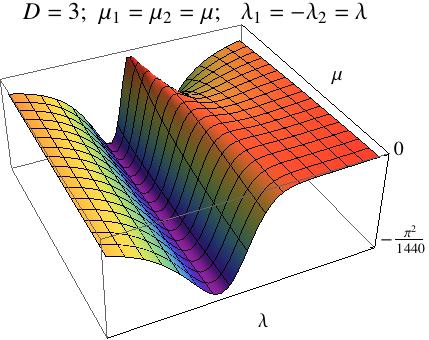}}
\caption{\footnotesize{Contour plot (left) and 3D plot (right) of $E_0^{{\rm int}} (2a)^3/\Sigma_3$ when the $\delta^\prime$ couplings of the two plates are identical in modulus but have different signs. We thus deal with \lq\lq opposite objects\rq\rq and find accordingly that the vacuum energy per unit surface is negative throughout the whole $\mu-\lambda$ quadrant. The minimum energy density is $-\pi^2/1440$ and occurs either at the straight line $\lambda=-2$, i.e., for Dirichlet boundary conditions, or, at the point $(\mu=0,\lambda=2)$, an special case of Robin boundary conditions. It is interesting to compare these Figures with Figure 3(right) and Figure 5 showing vacuum energies in the $1D$ case.}}
\label{3d-opp}
\end{figure}
\end{center}
\begin{center}
\begin{figure}[htpb]
\center{\includegraphics[width=8.55cm]{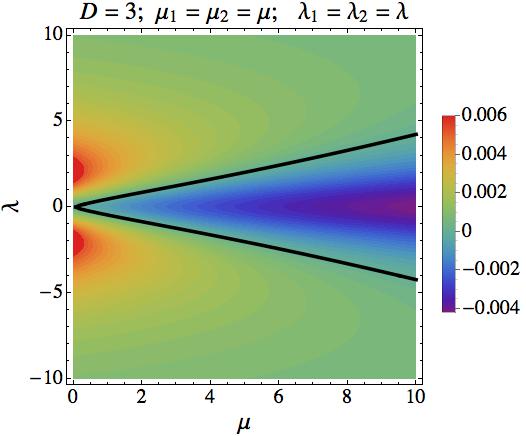}\includegraphics[width=8.25cm]{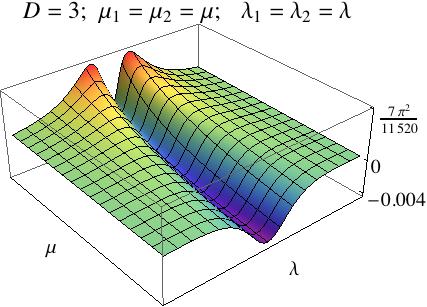}}
\caption{\footnotesize{Contour plot (left) and 3D plot (right) of $E_0^{{\rm int}} (2a)^3/\Sigma_3$ when both plates are identical. The vacuum energy may be positive, repulsive quantum vacuum force, negative, attractive quantum vacuum force, or zero, null quantum vacuum force, in different zones of the positive $\mu-\lambda$ quadrant. The thick black curve in the contour plot represents the zero energy density curve, whereas the maximum energy density is $7\pi^2/11520$ and occurs at  the two points: $(\mu=0,\lambda=\pm 2)$. The minimum energy density in this case is again  $-\frac{\pi^2}{1440}$ and occurs at the infinity point $(\mu=+\infty,\lambda=0)$,  the other regime where the field  fluctuations feel Dirichlet boundary conditions at both plates, see Reference \cite{Munoz-Castaneda:2013yga}. }}
\label{3d-iden}
\end{figure}
\end{center}
\end{widetext}

Regarding the dialectic \lq\lq opposite\rq\rq versus \lq\lq identical\rq\rq we have collected evidence about the fact that similar patterns are followed by the vacuum energy density in higher dimensions as compared with the one-dimensional situation.
The surge of repulsive, null, and attractive quantum vacuum forces between identical $1D$ plates, whereas opposite plates only suffer attraction, are properties also happening in higher dimensions. The theorem by Kenneth and Klich remains valid when the plates present a $\mathbb{Z}^2$ symmetry with respect to the hyperplane equidistant from both plates, i.e., when $\lambda_1=-\lambda_2$ and $\mu_1=\mu_2$, see Figure \ref{3d-opp}. If both plates are identical, i.e., $\lambda_1=\lambda_2$ and $\mu_1=\mu_2$, the $\mathbb{Z}^2$ symmetry is lost and therefore the quantum vacuum interaction energy between plates can be positive , negative, or zero, giving rise respectively to repulsive, attractive, or null quantum vacuum forces, as one can see in Figure \ref{3d-iden}.

\section{Conclusions and further comments}\label{conclusions}
In sum we draw the following conclusions from this work:
\begin{itemize}
\item We delved further into the physical aspects of the $\mu \delta+\lambda\delta^\prime$ point interaction relying on the natural definition of this potential by Gadella {\it et al} \cite{Gadella20091310} as a self-adjoint extension of the free particle Hamiltonian. In particular, we  
explained the special physical features of the scattering amplitudes and the bound state wave function at $\lambda=\pm 2$. For $\lambda=\pm2$ the scattering is completely opaque, with no transmission at all, and the probability amplitude of the particle in the bound state is restricted to only one side of the point where the point potential is placed.

\item A novel development achieved in this paper is the solution of the spectral problem of two pairs of $\delta-\delta^\prime$ point interactions. 
We have identified the scattering data of the quantum Hamiltonian defined by imposing over the eigenfunctions the Kurasov-Gadella et al. matching conditions at the two points on the real line where the interactions sit. The scattering amplitudes are accordingly analytically determined.
More difficulties arise in elucidating the energies of the bound states from the poles of the $S$-matrix on the positive imaginary half-axis in the momentum complex plane because it requires solving a transcendent equation. Nevertheless, the number of these poles (0,1, or, 2) has been 
identified in some significative two-dimensional sub-zones of the space of couplings.

\item The spectral data of the two $\delta-\delta^\prime$ Hamiltonian have been used as the one-particle states of one scalar quantum field theory in $(1+1)$-dimensional Minkowski space-time. By considering this point of view we have demonstrated that plates mimicked by $\delta-\delta^\prime$ potentials can be interpreted as a generalisation of Robin boundary conditions. When the $\delta^\prime$ coupling in the left is either set to $\lambda_1=2$ or the coupling in the right is fixed to $\lambda_2=-2$ the quantum fluctuations in the interval $[-a,a]$ coresponding to scattering waves incoming either from the left or from the right satisfy Robin boundary conditions provided that restrictions to a discrete set of momenta complying with a certain spectral conditions are imposed. If the arrangement $\lambda_1=2$, $\lambda_2=-2$, is chosen both the left and right movers satisfy Robin boundary conditions under a third, stronger, spectral condition. Thus, the QFT generated between plates becomes unitary and is identified with a massless scalar QFT defined over the interval with the quantum fields satisfying Robin boundary conditions. 
Therefore because one $\delta-\delta^\prime$ potential is understood as the idealization of a plate in a Casimir setup we analytically mimicked all the plates compatible with Robin boundary conditions. We should mention at this point that the opposite arrangement, $\lambda_1=-2$, $\lambda_2=2$,
gives rise to an unitary QFT of massless scalar fields complying with Dirichlet boundary conditions.

\item The main results reflect the quantum vacuum interaction energies arising between two 1D plates idealized as two $\delta-\delta^\prime$ point potentials. The calculations have been performed in the $TGTG$ formalism of Kenneth and Klich. We thus deal with two points, rather than two compact objects. Numerical evaluations of the integral in the $TGTG$ formula show that the quantum vacuum forces between two $\delta-\delta^\prime$ point plates are attractive, repulsive or null depending on the parameter space zones where the calculation is carried out. We avoided the parameter space zones where bound states exist and the theory ceases to be unitary leaving room to absorption/emission processes.

\item We have been able to obtain an analytical expression of the quantum vacuum interaction energy between two pure $\delta^\prime$ walls. In this case the quantum vacuum interaction is also positive (repulsive Casimir force), negative (attractive Casimir force), and zero (null Casimir force). 

\item Calculations in higher dimensional spaces of quantum vacuum energies per unit surface have also been achieved in the framework of the ${\rm TGTG}$ paradigm. In these cases the plates are $D-1$-dimensional hyper-cubes. We found similar patterns in three and one dimensions followed by the dependence of the vacuum energies on the couplings although the scale of the energies induced by vacuum fluctuations changes.

\item If the two $\delta-\delta^\prime$ pairs of couplings are identical,  $\mu_1=\mu_2$, $\lambda_1=\lambda_2$,  our calculations show that the quantum vacuum force between these two infinitely thin plates is repulsive. Apparently this behaviour does not fit within the framework of the Kenneth-Klich theorem, \lq\lq Opposites attract\rq\rq, see Ref. \cite{kenneth06}. The basic assumption in the proof of the Kenneth-Klich theorem is the characterization of \lq\lq opposite\rq\rq objects as two identical bodies placed in a $\mathbb{Z}_2$ reflection symmetrical way with respect to the hyperplane equidistant to the two objects. This mirror symmetry does not hold when the potentials mimicking the objects are not parity invariant, as happens with the $\delta-\delta^\prime$ plates. Understanding the $\delta^\prime$ coupling $\lambda$ as the polarizability perpendicular to the plate two identical plates symmetrical with respect to the hyperplane centred at the middle point are not opposite: the polarizabilities point in the same direction. In this case the two objects are opposite if the $\delta$ couplings are identical, $\mu_1=\mu_2$, 
but the $\delta^\prime$ couplings differ in sign: $\lambda_1=-\lambda_2$. In this arrangement the double $\delta-\delta^\prime$ potential is parity invariant and our results fit perfectly with the Kenneth-Klich theorem  provided that this refinement of the concept of \lq\lq opposite\rq\rq is assumed. According to our calculations concerning the double $\delta-\delta^\prime$ set of plates, we conclude that, \lq\lq opposites\rq\rq always attract  but \lq\lq identical\rq\rq may repel, attract, or do not interact.

\item It is tempting to replace the $\delta^\prime$ interaction by a singular potential of the form: $\lambda {\rm sign}(x)\delta^\prime(x)$, i.e.,
a multiplicative $\delta^\prime/step$ potential similar to the additive $\delta/step$ potential treated, for instance, in Reference \cite{susy-delta}. By considering this \lq\lq even\rq\rq 
potential we would have $r_R(k)=r_L(k)$ because parity is preserved and the refinement of the concept of opposite objects adopted in the paper would be authomatic.The general procedure could be developeded along the same lines starting from the appropriate modification of the matching conditions.

\end{itemize}

\section*{Acknowledgements}
We are grateful for the fruitful discussions and comments from M. Bordag, M. Asorey, K. Kirsten, D. Vassilevich, I. Cavero-Pelaez, M. Gadella, L. M. Nieto, and G. Marmo. Also the authors acknowledge the hospitality of the Pedro Pascual Benasque Center for Science, where part of this work was done. This paper has been partially supported by the DFG project BO1112-18/1

\appendix
\section{$\delta-\delta^\prime$ interactions and boundary conditions}\label{append}


Recall the matching conditions (\ref{1ddp-matchcond1},\ref{1ddp-matchcond2}) that define the $\delta-\delta^\prime$ point interaction:
\begin{eqnarray*}
&&\psi_k(0\uparrow)=\frac{1+\lambda/2}{1-\lambda/2}\psi_k(0\downarrow)  \\
&&\psi_k^\prime(0\uparrow)=\frac{1-\lambda/2}{1+\lambda/2}\psi_k^\prime(0\downarrow) +\frac{\mu}{1-\lambda^2/4}\psi_k(0\downarrow) .
\end{eqnarray*}
Our goal in this Appendix is to understand the $\delta-\delta^\prime$ interaction as a physical implementation of Robin boundary conditions. With this aim in mind we write the matching conditions (\ref{1ddp-matchcond1}-\ref{1ddp-matchcond2}) as a selfadjoint extension of the free particle Hamiltonian in the Asorey-Munoz-Castaneda formalism, see Reference \cite{mc-asorey}. We thus rewrite the $\delta-\delta^\prime$ matching conditions in the form
\begin{eqnarray}
&& \Psi_+={\bf U}_{\delta\delta^\prime}\Psi_-,\label{qbc}\\ &&
\Psi_\pm=\left(\begin{array}{c}\vspace{0.12cm}\mu \psi(0\uparrow) \pm i\psi^\prime(0\uparrow)\\ \vspace{0.12cm}\mu \psi(0\downarrow) \mp i\psi^\prime(0\downarrow)\end{array}\right)=\left(\begin{array}{c}\vspace{0.12cm}\Psi_\pm^{(1)}\\ \vspace{0.12cm}\Psi_\pm^{(2)}\end{array}\right),
\nonumber
\end{eqnarray}
where ${\bf U}_{\delta\delta^\prime}$ is a $2\times 2$ unitary matrix. 

Because 
\begin{eqnarray*}
\left(\begin{array}{c}\vspace{0.12cm}\Psi_+^{(1)}\\ \vspace{0.12cm}\Psi_-^{(1)}\end{array}\right)&=&
\left(
\begin{array}{cc}
 1 & i \\
 1 & -i \\
\end{array}
\right)\cdot
\left(\begin{array}{c}\vspace{0.12cm}\mu \psi(0\uparrow)\\ \vspace{0.12cm}\psi^\prime(0\uparrow)\end{array}\right)
\\
\left(\begin{array}{c}\vspace{0.12cm}\Psi_+^{(2)}\\ \vspace{0.12cm}\Psi_-^{(2)}\end{array}\right)&=&
\left(
\begin{array}{cc}
 1 & -i \\
 1 & i \\
\end{array}
\right)\cdot
\left(\begin{array}{c}\vspace{0.12cm}\mu \psi(0\downarrow)\\ \vspace{0.12cm}\psi^\prime(0\downarrow)\end{array}\right)
\end{eqnarray*}
a reshuffling of the matching conditions (\ref{1ddp-matchcond1}-\ref{1ddp-matchcond2}) in the form
\begin{equation}
\left(\begin{array}{c}\vspace{0.12cm}\Psi_+^{(1)}\\ \vspace{0.12cm}\Psi_-^{(1)}\end{array}\right)={\bf W}_{\delta\delta^\prime}\left(\begin{array}{c}\vspace{0.12cm}\Psi_+^{(2)}\\ \vspace{0.12cm}\Psi_-^{(2)}\end{array}\right)\label{sys-w}
\end{equation}
demands that the new matrix ${\bf W}_{\delta\delta^\prime}$ is a similarity transformation of the old matrix:
\begin{eqnarray}
{\bf W}_{\delta\delta^\prime}&=&\left(
\begin{array}{cc}
 1 & i \\
 1 & -i \\
\end{array}
\right)\cdot
\left(
\begin{array}{cc}
 \frac{1+\lambda/2}{1-\lambda/2} & 0 \\
 \frac{1}{1-\lambda^2/4} & \frac{1-\lambda/2}{1+\lambda/2} \\
\end{array}
\right)\cdot
\left(
\begin{array}{cc}
 1 & -i \\
 1 & i \\
\end{array}
\right)^{-1}\nonumber \\
&=&\frac{1}{4-\lambda^2}\left(
\begin{array}{cc}
 4\lambda+2i & 4+\lambda^2+2i \\
 4+\lambda^2-2i & 4\lambda-2i \\
\end{array}
\right)\nonumber
\end{eqnarray}
The second equation in the linear system (\ref{sys-w}) can be recast as
\begin{equation}
\Psi_+^{(2)}=\frac{1}{\left({\bf W}_{\delta\delta^\prime}\right)_{2,1}}\left(\Psi_-^{(1)} -\left({\bf W}_{\delta\delta^\prime}\right)_{2,2}\Psi_-^{(2)}\right),
\end{equation}
and used in the first equation of (\ref{sys-w}) to remove the dependence in $\Psi_+^{(2)}$:
\begin{equation}
\Psi_+^{(1)}=\frac{\left({\bf W}_{\delta\delta^\prime}\right)_{1,1}\Psi_-^{(1)} -\det\left({\bf W}_{\delta\delta^\prime}\right)\Psi_-^{(2)}}{\left({\bf W}_{\delta\delta^\prime}\right)_{2,1}}
\end{equation}
These two equations above identify the Asorey-Munoz-Castaneda quantum boundary condition (\ref{qbc}) coming from the matching conditions (\ref{1ddp-matchcond1}-\ref{1ddp-matchcond2}) defining the $\delta-\delta^\prime$ point interaction. The corresponding matrix is
\begin{eqnarray}
\hspace{-0.3cm}{\bf U}_{\delta\delta^\prime}&=&\frac{1}{\left({\bf W}_{\delta\delta^\prime}\right)_{2,1}}\left(
\begin{array}{cc}
 \left({\bf W}_{\delta\delta^\prime}\right)_{1,1} & -\det{\bf W}_{\delta\delta^\prime} \\
 1 &-\left({\bf W}_{\delta\delta^\prime}\right)_{2,2} \\
\end{array}
\right)\nonumber \\
&=&\frac{1}{4+\lambda^2-2i}\left(
\begin{array}{cc}
 4\lambda+2i & 4-\lambda^2 \\
 4-\lambda^2 & -4\lambda+2i \\
\end{array}
\right).\label{uddp}
\end{eqnarray}
It is straightforward to check that ${\bf U}_{\delta\delta^\prime}$ is indeed a unitary matrix, and therefore it defines a selfadjoint extension
of the free particle Hamiltonian.

\subsection{The $\lambda\rightarrow\pm 2$ limits of the $\delta^\prime$ coupling and Robin boundary conditions}
The $\lambda\rightarrow \pm 2$ limits of the unitary matrix (\ref{uddp}) are
\begin{equation}
\lim_{\lambda\rightarrow2}{\bf U}_{\delta\delta^\prime}=\left(\begin{array}{cc}
 \frac{4-i}{4+i} & 0 \\
 0 & -1 \\
\end{array}
\right) \, \, , \, \, 
\lim_{\lambda\rightarrow-2}{\bf U}_{\delta\delta^\prime}=\left(\begin{array}{cc}
 -1 & 0 \\
 0 & \frac{4-i}{4+i} \\
\end{array}
\right) \nonumber
\end{equation}
We remark that, see Reference \cite{mc-asorey}:
\begin{itemize}
\item $\psi(a)\pm i a \psi^\prime (a)=-(\psi(a)\mp i a \psi^\prime (a))$ requires that $\psi(a)=0$, i.e., it is equivalent to imposing a Dirichlet boundary condition at $x=a$.
\item $\psi(a)\pm ia \psi^\prime (a)=e^{2 i\alpha}(\psi(a)\mp i a \psi^\prime (a))$, or $\psi(a)\tan(\alpha)\mp a \psi^\prime(a)=0$, is equivalent to demanding a Robin boundary condition characterized by the angle $\alpha$ at $x=a$.
\end{itemize}
We find that:
\begin{enumerate}
\item For a $\lambda=2$ $\delta^\prime$ coupling 
\[
{\rm (a)} \, \, \psi(0\uparrow)-\frac{4}{\mu}\psi^\prime(0\uparrow)=0 \quad , \quad {\rm (b)} \, \, \psi(0\downarrow)=0
\]
The pair of data $\{\mu \psi(0\uparrow),\psi^\prime(0\uparrow)\}$ subjected to the matching conditions (\ref{1ddp-matchcond1}-\ref{1ddp-matchcond2}) satisfies Robin boundary conditions at the origin if $\lambda=2$ and $\tan\alpha=\frac{\mu a}{4}$. The second pair $\{\mu \psi(0\downarrow),\psi^\prime(0\downarrow)\}$, however, satisfies Dirichlet boundary conditions when $x=0$ and $\lambda=2$.
\item If $\lambda=-2$ the two pairs of data exchange their r$\hat{\rm o}$les: 
\[
{\rm (a)} \, \,  \psi(0\uparrow)=0 \quad , \quad {\rm (b)} \, \, \psi(0\downarrow)+\frac{4}{\mu}\psi^\prime(0\downarrow)=0
\]
$\{\mu\psi(0\uparrow),\psi^\prime(0\uparrow)\}$ satisfies Dirichlet boundary conditions, whereas $\{\mu\psi(0\downarrow),\psi^\prime(0\downarrow)\}$ satisfies Robin boundary conditions.
\end{enumerate}
It is possible therefore to tune both $\delta-\delta^\prime$ interactions in the Casimir setup as Robin plates and obtain the same result as Ref. \cite{rom-jpa35}. This
choice requires $\lambda_1=2$ to be set in the plate at $x=-a$ and $\lambda_2=-2$ as the $\delta^\prime$-coupling at $x=a$. Under these circumstances
the quantum fluctuations between plates experience Robin boundary conditions. According to the analysis above this happens when the plate that is placed at $x=-a$, collecting the contribution from $r_L$ to the vacuum energy, is tuned at the $\lambda_1=2$ opaque point, whereas the plate that is placed at $x=a$, picking the contribution of $r_R$ to the vacuum energy, sets its $\delta^\prime$-coupling to be $\lambda_2=-2$.
\bibliography{2ddp-bibliogr}
\bibliographystyle{unsrt}



\end{document}